# Magnitude Of Earthquakes Controls The Size Distribution Of Their Triggered Events

**Shyam Nandan[1], Guy Ouillon[2], Didier Sornette[3]**

[1]ETH Zürich, Swiss Seismological Service, Sonneggstrasse 5, 8092 Zürich, Switzerland

[2]Lithophyse, 4 rue de l'Ancien Sénat, 06300 Nice, France

[3]ETH Zürich, Department of Management, Technology and Economics, Scheuchzerstrasse 7, 8092 Zürich, Switzerland

Corresponding author: Shyam Nandan (snandan@ethz.ch)

**Key Points:**

- Magnitudes of background earthquakes are distributed according to Gutenberg-Richter (GR) law.

- The magnitudes of earthquakes directly triggered by a given earthquake follow a kinked GR law with two branches.

- The location of the kink coincides with the magnitude of the triggering event.

## Abstract

The driving concept behind one of the most successful statistical forecasting models, the ETAS model, has been that the seismicity is driven by spontaneously occurring background earthquakes that cascade into multitudes of triggered earthquakes. In nearly all generalizations of the ETAS model, the magnitudes of the background and the triggered earthquakes are assumed to follow Gutenberg-Richter law with the same exponent ($\beta$-value). Furthermore, the magnitudes of the triggered earthquakes are always assumed to be independent of the magnitude of the triggering earthquake.

Using an EM algorithm applied to the Californian earthquake catalogue, we show that the distribution of earthquake magnitudes exhibits three distinct $\beta$-values: $\beta_b$ for background events; $\beta_a - \delta$ and $\beta_a + \delta$, respectively, for triggered events below and above the magnitude of the triggering earthquake; the two last values express a correlation between the magnitudes of triggered events with that of the triggering earthquake, a feature so far absent in all proposed operational generalizations of the ETAS model. The ETAS model incorporating this kinked magnitude distribution provides by far the best description of seismic catalogs and could thus have the best forecasting potential. We speculate that the kinked magnitude distribution may result from the system tending to restore the symmetry of the regional displacement gradient tensor that has been broken by the initiating event. The general emerging concept could be that while the background events occur primarily to accommodate the symmetric stress tensor at the boundaries of the system, the triggered earthquakes are quasi-Goldstone fluctuations of a self-organized critical deformation state.

## 1 Introduction

The driving concept behind one of the most successful statistical forecasting models, the Epidemic Type Aftershock Sequence (ETAS) model, has been that relatively rare and independent background events, powered by plate tectonics, cascade into multitudes of triggered events sharing the same space-time-magnitude distribution laws. This model relies on the master equation:

$$\lambda(t, x, y, m | \mathcal{H}_t) = \mu(x, y) f_b(m) + \sum_{i : t_i < t} g(t - t_i, x - x_i, y - y_i, m_i) f_a(m | m_i) \qquad (1)$$

which defines the rate $(\lambda)$ of earthquakes with magnitude $m$ at time $t$ and location $(x, y)$ conditioned on the history $(\mathcal{H}_t)$ of seismicity until $t$. $\mu(x, y)$ is the time-independent background rate of events (nucleating with a magnitude distribution $f_b(m)$), while the kernel $g$ defines the rate of events triggered by a previous shock $i$ as a function of the time difference and spatial distance between triggering and triggered events, with a conditional magnitude distribution $f_a(m | m_i)$. Hereafter, following one of the standard forms in the literature [Zhuang et al., 2002, 2004; Ogata, 1998], the general form of $g$ is assumed to be:

$$g(t - t_i, x - x_i, y - y_i, m_i) =$$
$$K e^{a(m_i - M_0)} \frac{\omega c^\omega}{\{t - t_i + c\}^{1+\omega}} \frac{\rho d^\rho e^{\gamma \rho (m_i - M_0)}}{\pi \{(x - x_i)^2 + (y - y_i)^2 + d e^{\gamma (m_i - M_0)}\}^{1+\rho}} \qquad (2)$$

The standard ETAS model (see Text S1) assumes that [Helmstetter and Sornette, 2002]: (i) the magnitudes of background and triggered earthquakes (in number proportional to $e^{a(m_i - M_0)}$, where $a$ is the productivity exponent) are distributed according to the same GR law, $f_b(m) = f_a(m | m_i) \sim \beta e^{-\beta(m - M_0)}$, $M_0$ being the minimum magnitude of an event able to trigger some others, so that both background and triggered events differ only in their space and time rates, not on their respective physical origins ; (ii) the magnitudes of the triggered earthquakes and their distribution does not depend on the magnitude of the earthquake that triggered them, so that the modelled system has no direct memory of the size of past events (it only has an indirect one since large events have an abundant progeny and therefore a significant influence on the future seismicity rate [Helmstetter and Sornette, 2003])**.** However, recent studies have presented empirical evidence against the above assumptions. For instance, a stochastic declustering of the Japanese Meteorological Agency (JMA) catalogue (using the standard ETAS model [Ogata, 1988]) shows that the exponent $\beta$ of the GR law of the direct aftershocks decreases with the mainshock magnitude, implying that large mainshocks tend to trigger larger direct aftershocks [Zhuang et al., 2004]. It is important to note that, although the ETAS model makes no distinction between earthquakes in terms of "mainshocks", "aftershocks" and "foreshocks" like many other seismicity modelling approaches [see for instance, Zalliapin and Ben-Zion, 2013; 2018; Reasenberg, 1985], for brevity, we refer to triggering and triggered earthquakes throughout the paper as mainshocks and aftershocks. Zhuang et al. [2004] also found that the GR laws of the background and all the triggered earthquakes display different exponents $\beta$. Applying a non-parametric stochastic declustering algorithm [Marsan and Lengliné, 2008] to the global CMT catalogue, Nichols and Schoenberg [2014] showed that the average magnitude of the directly triggered earthquakes tends to systematically increase with their mainshock magnitude. Similar observations for the Italian as well as the South Californian catalogues of earthquakes have been presented [Spassiani and Sebastiani, 2016]. Indeed, for a given set of events, the maximum likelihood estimator for $\beta$ is inversely proportional to their average magnitude [Aki, 1965], so that the aforementioned works all reach the same conclusion.

These observations motivate the use of the generalised Vere-Jones ETAS (gV-ETAS) model (inspired from [Vere-Jones, 2005; Saichev and Sornette, 2005]), which assumes that:

$$f_b(m) = \beta_b e^{-\beta_b(m-M_0)} \tag{3}$$

$$f_a(m|m_i) = \begin{cases} C_1 e^{-(\beta_a+\delta)m} & \forall\, m > m_i \\ C_2 e^{-(\beta_a-\delta)m} & \forall\, m \le m_i \end{cases} \tag{4}$$

where $C_1$ and $C_2$ are such that $f_a(m)$ is continuous and $\int_{M_0}^{\infty} f_a(m)dm = 1$. Applying the continuity and normalization constraints on the conditional magnitude distribution of the direct aftershocks in Equation (4), we find that $C_2 = \left[ \frac{2\delta}{\beta_a^2 - \delta^2} \left\{ \left( \frac{\beta_a+\delta}{2\delta} \right) e^{-(\beta_a-\delta)M_0} - e^{-(\beta_a-\delta)m_i} \right\} \right]^{-1}$ and $C_1 = C_2 e^{2\delta m_i}$.

The most interesting aspect is that the magnitude distribution of the triggered earthquakes is such that their magnitudes tend to cluster around that of their parent earthquake. The original models [Vere-Jones, 2005; Saichev and Sornette, 2005] were motivated by the search for the minimal model that achieves the condition of exact statistical self-similarity, which imposes $a = \beta_a = \beta_b$. Exact statistical self-similarity means that the only scale relevant to the production of future aftershocks (over all generations) is the size of their ancestor. In this self-similar version [Vere-Jones, 2005], the gV-ETAS does not require a minimum triggering magnitude $M_0$ nor a maximum magnitude. In contrast, the standard ETAS model and the gV-ETAS model with $a < \beta_a \ne \beta_b$ require such an "ultra-violet" magnitude cut-off $M_0$ to prevent the cloud of very small earthquakes to overwhelm the production of events and lead to a diverging seismicity rate [Saichev and Sornette, 2005; Sornette and Werner, 2005].

Despite its mathematical elegance, no empirical evidence has yet been provided to test the validity of this model. Here, we extend an expectation maximisation (EM) scheme [Veen and Schoenberg, 2008] to calibrate the gV-ETAS model on the Californian earthquake catalogue directly, and then compare the performance of several versions of gV-ETAS to that of the standard ETAS model using standard likelihood ratio tests. We also perform rigorous synthetic tests to eliminate sources of biases that could potentially distort our findings. We then discuss the results within the framework of spontaneous symmetry breaking and restoration at the onset of self-organised critical transitions.

## 2 Method

### 2.1 Five variants of the gV-ETAS model

In this study, we have considered the following five variants of the gV-ETAS model:

1. *Model 1 (standard ETAS or gV-ETAS model with $\delta = 0$ and $\beta_b = \beta_a$):* The magnitudes of the background and the directly triggered (DT) earthquakes both follow a pure GR law with the same exponent $\beta$.

2. *Model 2 (gV-ETAS model with $\delta = 0$):* The magnitudes of the background and the DT earthquakes are distributed according to pure GR laws, but with different exponents $\beta_b$ and $\beta_a$, respectively. Model 2 does not assume any magnitude correlation, just that aftershocks and mainshocks may stem from different physical processes.

3. *Model 3 (Vere-Jones model):* The magnitudes of the background earthquakes are distributed according to the GR law with exponent $\beta_b$, whereas the magnitudes of the DT earthquakes are distributed according to the GR law with exponents $\beta_b - \delta$ ($\forall\, m \leq m_i$) and $\beta_b + \delta$ ($\forall\, m > m_i$), where $m_i$ represents the mainshock magnitude.

4. *Model 4 (special gV-ETAS model):* The magnitudes of the background earthquakes are distributed according to the GR law with exponent $\beta_b = \beta_a + \delta$, whereas the magnitudes of the DT earthquakes are distributed according to the GR law with exponents $\beta_a - \delta$ ($\forall\, m \leq m_i$) and $\beta_a + \delta$ ($\forall\, m > m_i$), where $m_i$ represents the mainshock magnitude. The underlying idea is that earthquakes which appear as background events could (in an extreme case) have been all triggered by events smaller than the magnitude completeness threshold, and thus could be 'large' aftershocks of those unobserved events. We thus force them to belong to the $\beta + \delta$ branch of the magnitude distribution.

5.  *Model 5 (gV-ETAS model):* The magnitudes of the background earthquakes are distributed according to the GR law with exponent $\beta_b$, whereas the magnitudes of the DT earthquakes are distributed according to the GR law with exponents $\beta_a - \delta$ ($\forall \, m \leq m_i$) and $\beta_a + \delta$ ($\forall \, m > m_i$), where $m_i$ represents the mainshock magnitude. This is the most general model featuring three independent parameters.

Note that models 1 to 4 are nested within Model 5, while Model 1 is nested within models 2 to 5.

## 2.2 Estimating the parameters of the gV-ETAS model using the EM algorithm

To estimate the parameters, $\theta = \{\mu, K, a, c, \omega, d, \gamma, \rho, \beta_b, \beta_a, \delta\}$, of the gV-ETAS model defined in the previous section, we extend the EM algorithm proposed by [Veen and Schoenberg, 2008]. As described in [Nandan et al., 2017], the algorithm can be broken down into two steps:

1.  *Expectation step (or E-step):* Given the guess/estimate of the parameters at the $k^{th}$ step, $\hat{\theta}^k$, we first compute the triggering probability, $P_{ij}^k$, according to the following equation:

$$P_{ij}^k = \frac{g(t_j - t_i, x_j - x_i, y_j - y_i, m_i | \hat{\theta}^k) f_a(m_j | m_i, \hat{\theta}^k)}{\mu f_b(m_j | \hat{\theta}^k) + \Sigma_{i \, \forall \, t_i < t_j} g(t_j - t_i, x_j - x_i, y_j - y_i, m_i | \hat{\theta}^k) f_a(m_j | m_i, \hat{\theta}^k)} \qquad (5)$$

$P_{ij}^k$ quantifies the probability that the $j^{th}$ earthquake in the catalog has been triggered by the $i^{th}$ earthquake ($\forall \, t_i < t_j$), i.e. the probabilistic genealogy tree of earthquakes present in the catalog. Using $P_{ij}^k$, we can then estimate: (a) the independence probability of the $j^{th}$ earthquake, $IP_j^k = 1 - \sum_{i \, \forall t_i < t_j} P_{ij}^k$, i.e. the probability that the $j^{th}$ earthquake in the catalog has not been triggered by any earthquake preceding it and is thus a background earthquake; (b) the number of direct aftershocks of the $i^{th}$ earthquake, $\psi_i^k = \sum_{j \, \forall t_i < t_j} P_{ij}^k$, and (c) the total number of background earthquakes $\phi^k = \sum_j IP_j^k$.

2. *Maximization step (or M-step):* In this step, we maximise the expected complete data log likelihood, $l^k(\theta)$, with respect to the parameters $\theta$. $l^k(\theta)$ is defined according to the following equation:

$$l^k(\theta) = -\log\Big(\Gamma(\phi^k + 1)\Big) - \mu AT + \phi^k \log(\mu AT) +$$
$$\sum_i \Big\{ -\log\Big(\Gamma(\psi_i^k + 1)\Big) - G_i(\theta) + \psi_i^k \log(G_i(\theta)) \Big\} +$$
$$\sum_{ij} P_{ij}^k log\left\{\frac{g(t_j - t_i, x_j - x_i, y_j - y_i, m_i)}{G_i(\theta)}\right\} + LL_q^{\Delta m,k} \qquad (6)$$

In the above equation, $A$ and $T$ respectively represent the area of the study region (in $km^2$) and the total time span of the catalog, $T_2 - T_1$ (in days). $G_i(\theta)$ is the expected number of offsprings triggered by an earthquake $(t_i, x_i, y_i, m_i)$ within the study region $S$ and the time period $[T_1, T_2]$ and is given by $\int_{\max(t_i,T_1)}^{T_2} \iint_S g(t - t_i, x - x_i, y - y_i, m_i) \, dx \, dy \, dt$. Unlike [Veen and Schoenberg, 2008; Nandan et al., 2017], the magnitude distribution of the background earthquakes and the conditional magnitude distribution of the aftershocks also contribute to the expected complete data log likelihood, according to the log likelihood $LL_q^{\Delta m,k}$, where the subscript $q$ refers to the index of the gV-ETAS models defined in Section 2.1. These additional log-likelihood terms for the five variants of gV-ETAS models compared in this study have been explicitly defined in supplementary Text S2. The maximization of $l^k(\theta)$ gives the new estimates of the parameters, $\hat{\theta}^{k+1}$, which can be used as the input for the E-step. The E and M steps defined above are repeated as long as $|l^{k+1} - l^k| > 10^{-4}$.

## 3 Dataset and some necessary considerations

For the calibration of the five gV-ETAS models, we use the earthquake catalogue obtained from the Advanced National Seismic System (ANSS) database, including earthquakes in and around the state of California. More specifically, we use the collection polygon defined for the Regional Earthquake Likelihood Model (RELM) experiment [Schorlemmer and Gerstenberger, 2007] as the spatial boundary of the catalogue. All earthquakes (24,972) with $M \geq 3$, and depth $\leq 40\ km$ that occurred in the time period from January 1, 1981 to May 31, 2017, and enclosed in the above-mentioned spatial polygon, are used for the calibration of the stochastic models used in this study. The choice of $M \geq 3$ earthquakes has been justified by estimating independent temporal and spatial variation of magnitudes of completeness in the study region [Nandan et al., 2017]. Notice that we also performed tests to estimate the effect of the short-term incompleteness of the catalogue after large magnitude events (see Section 4.5).

An important consideration before the calibration of such stochastic models is that some of the parents/offspring of the earthquakes are missing in the dataset due to the finite space-time-magnitude limits imposed on the catalogue. While space and time limits are within the control of the modeller, the magnitude limit is primarily dictated by the magnitude of completeness of the catalogue governed by the space-time dependent sensitivity of the seismic networks. If unaccounted for, these anthropogenic limits can lead to biases in the estimates of the parameters, as has been extensively demonstrated in the literature [Seif et al., 2017; Wang et al. 2010; Schoenberg et al., 2010; Harte, 2015; Helmstetter et al., 2006; Werner et al., 2011; Hainzl et al., 2013; Omi et al., 2014; Kagan, 2004; Hainzl, 2016]. We use "auxiliary catalogues" to limit the influence of the missing parents on the genealogy tree of earthquakes derived from the calibration of the stochastic models. The earthquakes in the auxiliary catalogue can act only as parents and contribute in constraining the triggering function. The two types of auxiliary earthquakes that are used in this study are: (i) all earthquakes that lie within the "Auxiliary polygon" (a polygon concentric to the aforementioned collection polygon with $+1°$ larger radius), occurred between January 1, 1971 and December 31, 1980 and have $M \geq 3$; (ii) earthquakes that lie in the area between the auxiliary polygon and the collection polygon, that occurred between January 1, 1981 and May 31, 2017 and have $M \geq 3$. The total contribution of these two types of auxiliary earthquakes can be easily evaluated by integrating the triggering function with appropriate: (a) time limits, [January 1, 1981; May 31, 2017] for the first type of auxiliary earthquakes, and $[t_i, \text{May } 31, 2017]$ for the second type of auxiliary earthquakes, and (b) space limits, the area enclosed by the collection polygon.

## 4 Results

### 4.1 Comparing the performance of the gV-ETAS and classical ETAS models using Wilks' test

Table 1 lists the value of the parameters of all the five competing gV-ETAS models, which have been obtained by maximising the expected data log-likelihood (see Section 2.2).

Based on the likelihood scores obtained for each model, we compute the pairwise likelihood ratio test statistics (Wilks' test) for all the nested hypotheses. The log-likelihood ratio test allows us to quantify how much more likely is the observed data under one model (say, Model 5 with three free parameters) than under another model (say, Model 1 with one free parameter), accounting for the extra degree(s) of freedom that the complex model enjoys relative to its simpler counterpart. Given two models (such as models 5 and 1), the Wilks test assumes that the simpler model is correct and, under this null hypothesis, gives the probability $p_S^C$ (where the superscript C refers to the complex model and subscript S refers to the simpler model) that this is the case for the given data set when compared with the more complex model. $p_S^C$ is calculated by entering twice the difference of the log-likelihoods of the two models as the argument of the chi-squared distribution with a number $df$ of degrees of freedom equal to the difference in the number of free parameters of the complex model and the simpler one. It has been suggested in the literature [Pinheiro and Bates, 2000] that this way of using chi-squared distribution to compute the Wilks' test statistics is rather naïve, as the difference in the degrees of freedom between the complex and the simpler model is not simply the difference in the number of non-redundant parameters of the two. Furthermore, it has been shown [Stram and Lee, 1994; Self and Liang, 1987] that the Wilks' statistics obtained in the naïve way is conservative: the corrected test statistics [Chernoff, 1954] being smaller than the naïve statistics. Nevertheless, in the following, we use the naïve approach, as it is easily implementable and a rejection using the naïve Wilks' statistics would guarantee a rejection using the corrected Wilks' statistics.

We obtain the corresponding p-values: $p_{H_1}^{H_2} = 6.3 \times 10^{-5}$, $p_{H_1}^{H_3} = 5.1 \times 10^{-279}$, $p_{H_1}^{H_4} = 4.9 \times 10^{-249}$, $p_{H_1}^{H_5} = 1.9 \times 10^{-283}$, $p_{H_2}^{H_5} = 1.2 \times 10^{-281}$, $p_{H_3}^{H_5} = 1.2 \times 10^{-7}$, $p_{H_4}^{H_5} = 5.5 \times 10^{-38}$ (the smaller $p_{H_j}^{H_i}$, the larger the probability that model $H_i$ holds vs. $H_j$). These extremely small p-values indicate that (a) model 5 is very significantly better than all the competing models; (b) models 2, 3 and 4 are significantly better than Model 1 (classical ETAS).

We show in the upper panels of Figure 1 the goodness of the fits between the theoretical (see Table 1) and the empirical (derived from the branching structure provided by each of the five variants of the gV-ETAS models) magnitude distributions of the triggered earthquakes conditioned on the mainshock magnitude for each of the five models. Those plots are constructed using the branching structure evidenced by the corresponding gV-ETAS model and the following steps:

*Step 1: Magnitude binning.* All earthquake magnitudes are first sorted in different magnitude bins $B_{i=1,....43}$: [3:3.1], [3.1:3.2],...[7.2:7.3].

*Step 2: Distribution of the magnitude of events triggered by mainshocks in each magnitude bin.* For each integer $k=1,...,43$, we consider all events that are falling within bin $B_k$, as well as their respective offsprings (whatever their magnitude) and the associated triggering probabilities derived by the declustering procedure. Using the latter as weights, we can compute the total number of triggered events within each bin $B_{j=1,...,43}$ conditioned to the fact that the triggering event is in bin $B_k$.

*Step 3: Offsetting magnitude bins to the same reference value.* The latter distribution is then offset along the magnitude axis so that the central value of the bin corresponding to the mainshock is 0.

*Intermezzo:* Steps 1-3 are repeated for each mainshock magnitude bin so that we get 43 such offset direct aftershock magnitude distributions. For the smallest mainshock magnitude bin, [3:3.1], the relative magnitudes of the direct aftershocks cover the range [0:4.3], as 3 is the smallest magnitude of the considered earthquakes in the catalogue and 7.3 is the largest one. For the next mainshock magnitude range, [3.1:3.2], the range of the associated triggered shifted magnitude PDF is [-0.1:4.2], then [-0.2,4.1] for the next mainshock magnitude bin, [3.2:3.3], and so on, up to [-4.3:0] for the largest mainshock magnitude bin. All shifted PDFs integrate to 1 over their respective definition interval.

*Step 4:* we now combine those shifted PDFs to shape a master PDF of relative magnitudes of direct aftershocks irrespective of the mainshock magnitude bin, to extend the range of relative magnitudes to [-4.3:4.3] and to reduce the scatter present in the individual PDFs. We first consider the two first shifted triggered magnitude intervals, [0:4.3] and [-0.1:4.2], their common definition range [0:4.2], as well as their corresponding shifted triggered magnitude PDFs $P_1$ and $P_2$. Integrating both PDFs over their common interval gives values $S_1 = \int_0^{4.2} P_1 dm < 1$ and $S_2 = \int_0^{4.2} P_2 dm < 1$, respectively. We define a scaling factor $R_{1,2} = S_1/S_2$, and from now on replace the full PDF $P_2$ by $R_{1,2} P_2$. Note that the new $P_2$ now does not integrate to 1 anymore over its full definition interval. We then use the same procedure to rescale $P_3$, which holds over the interval [-0.2:4.1], by computing a rescaling factor $R_{2,3} = S_2/S_3$ (the common integration interval for $S_2(P_2)$ and $S_3(P_3)$ is now [-0.1:4.1]). This rescaling step is sequentially repeated up to $P_{43}$, using the factor $R_{42,43}$. We are now in possession of 43 individual shifted density functions, which can be compared to each other, and are plotted as grey crosses on Figure 1. This master plot now spans the interval [-4.3:4.3], with a magnitude bin size of 0.1.

*Step 5:* Within each shifted triggered magnitude bin, we average the values of the grey crosses and get the black crosses shown on the same figures. The solid red lines in the upper panels show the theoretical PDFs of the direct aftershocks corresponding to each of the five tested hypotheses. These PDFs are obtained by using a mainshock magnitude set to 0 and the parameters reported in Table 1, ensuring that its integral over [-4.3:4.3] is the same as the one empirically obtained using the black crosses.

In the bottom panels of Figure 1, we demonstrate the goodness of fit between the empirical PDFs (black crosses) obtained from the independence probabilities estimated during the calibration of the five gV-ETAS models on the Californian catalogue and the theoretical PDF of background earthquakes with $\beta_b$ set to the values reported for each of the five models in Table 1.

In agreement with the pairwise *p*-values reported above, one can observe that the theoretical PDFs underlying models 3, 4 and 5 fit better the empirical PDF of the magnitudes of the triggered and the background earthquakes than do the theoretical PDFs of models 1 and 2. The improvement in the quality of the fits going from Model 3 to Model 4 or to Model 5 and from Model 1 to Model 2 is qualitatively less visible, but is quantitatively highly significant, as assessed rigorously via the Wilks' test.

## 4.2 How well do the five gV-ETAS models perform when the triggering probabilities are obtained from the classical ETAS model?

We first calibrate the classical ETAS model described by Equations (S1) and (S2) in Text S1 on the Californian catalog and obtain the following estimates of the parameters: $\mu = 2.31 \times 10^{-7} \, day^{-1}$, $K = 0.58, a = 1.00, c = 1.73 \times 10^{-2} \, day, \omega = 0.17, d = 0.24 \, km^2, \rho = 0.53$ and $\gamma = 1.11$. We also evaluate the triggering probability matrix $P_{ij}$ and the independence probability vector $IP_j$. These probabilities are then used to check for the self-consistency of the ETAS-based declustering, namely that the (output) branching properties should be identical to the (input) assumed ones: the magnitude distribution should be independent of the magnitude of the triggering events as well as of the triggering/triggered type of events.

Using the obtained $P_{ij}$ and $IP_j$, we then evaluate the performance of the magnitude distributions underlying the five variants of the gV-ETAS model. Table 2 compiles the maximum likelihood estimates of the parameters of magnitude distributions underlying the five models. The expressions of the maximum likelihood functions for the five models are given in the Equations S8 to S12 in Supplementary Text S2, as we test only the relevance of the various assumed shapes of the magnitude distributions.

As in Section 4.1, we quantify the likelihood of one model over the other using the standard pairwise log-likelihood ratio tests. We report the *p*-values ($p_S^c$ 's) of the pair-wise likelihood ratio tests in Table 2. Based on these *p*-values, we conclude that: (i) Model 5 (gV-ETAS) is highly significantly better than all other competing models; and (ii) models 2, 3 and 4 are significantly better than Model 1 (ETAS).

Figure 2 shows the goodness of the fits between the theoretical (obtained by fitting) and the empirical (derived from the $P_{ij}$ matrix provided by the standard ETAS model declustering algorithm) magnitude distributions of the triggered earthquakes conditioned on the mainshock magnitude for each of the five models. In each of the five top panels in the figure, the empirical PDF is the same, and obtained by first offsetting each mainshock magnitude ($m_i$) to 0 as well as its corresponding aftershocks, with magnitudes $m_j$, to relative magnitudes $m_{ij} = m_j - m_i$. We then estimate the distribution of the relative magnitudes $m_{ij}$, using the corresponding triggering probabilities $P_{ij}$, which are obtained from the calibration of the standard ETAS model on the Californian catalogue. Note that steps 1 to 5 used for constructing Figure 1 (Section 4.1) also apply to the construction of Figure 2, except that the triggering probabilities used for the construction of Figure 2 are always obtained from the calibration of the classical ETAS model on the data. The theoretical PDFs are computed using the maximum likelihood estimates of the parameters of models 1 to 5 listed in Table 2. In the bottom panels, we also show the correspondence between the empirical and theoretical PDFs of the magnitude distribution of the background earthquakes for each model. The empirical PDF is also the same in all panels and is obtained by estimating the distribution of magnitude, $m_j$, of each earthquake in the catalog using their independence probabilities $IP_j$ as weights. Solid red lines are the theoretical PDFs derived using the estimates of $\beta_a$ listed in Table 2.

In agreement with the *p*-values reported in Table 2, one can observe that the theoretical PDFs underlying models 3, 4 and 5 fit better the empirical PDF of the magnitudes of the triggered and the background earthquakes than do the theoretical PDFs of models 1 and 2. However, the visual improvement in the quality of the fits going from Model 3 to Model 4 or Model 5, and from Model 1 to Model 2 is quite small but has been assessed rigorously via the Wilks test.

It is important to stress that, in the preceding calibration exercise, we have estimated the triggering and independence probabilities $P_{ij}$ and $IP_j$ assuming that Model 1 (ETAS) is the true model (in order to determine the triggering probabilities $P_{ij}$). Our results demonstrate that the ETAS model is inconsistent, as is does not stand out as the best posterior fit (shown in Figure 2).

### 4.3 Does the Wilks test disproportionately favor the more complex gv-ETAS5?

It has been demonstrated in the literature that, on the one hand, the Wilks test tends to pick overly simple models when the sample size is small due to a lack of statistical power; on the other hand, it could disproportionately favor the more complex models when considering large sample sizes, which are ordained with high statistical power [Cudeck and Browne, 1983; Busemeyer and Wang, 2000]. Since we are dealing with large sample sizes (~ 25,000 earthquakes with $M \geq 3$ in the primary catalog), it is possible that the results of the Wilks test may have been disproportionately biased in favor of the more complex gvETAS5 model. To demonstrate that this is not the case, we perform the following test: we generate numerous (10,000) ~36 year-long synthetic earthquake catalogs within a square region with area $10^6 \; km^2$. The generating model for the each of the synthetic catalog is the gv-ETAS1 (i.e. the classical ETAS) model with the parameters: $\mu$ =92 earthquakes ($M \geq 3$) per year in the study region; $d = 0.25 \; km^2$; $\rho = 0.6$; $\gamma = 1.2$; $\omega = 0.2$; $c = 6.7 \times 10^{-3} days$; $K = 0.1$, $a = 2$ and $\beta_a = \beta_b = 2.3$. The magnitude of simulated earthquakes in the catalog are then binned at an interval of 0.1 to mimic the situation in the case of the real catalog. As the calibration of gv-ETAS models is computationally intensive, we restrict the size of the simulated catalogs in terms of number of earthquakes to 20,000-30,000. Note that this range contains the number of earthquakes present in the primary Californian catalog (~25,000) used in this study. For 10,000 such simulated catalogs, we calibrate the gv-ETAS1 and gv-ETAS5 models and compute their respective log-likelihood scores. Using the difference in the log-likelihood score and the difference in the number of free parameters of the two models, we compute the Wilks' statistics as in section 4.1 and 4.2 corresponding to each simulated catalog. We find that, in only 7.9% and 2.6% of the simulated catalogs, the Wilks' statistic is smaller than the standard rejection thresholds of 0.05 and 0.01, respectively, which would warrant a rejection of the gv-ETAS1 model in favor of gv-ETAS5 model as the true generating process. In absence of any bias, one would expect 5% and 1% of the simulated catalogs to obtain a Wilks statistics smaller than the rejection threshold of 0.05 and 0.01, respectively. The slightly larger values of 7.9% compared with 5% and 2.6% compared with 1% thus show a tendency for over-rejecting the simple model in favor of the more complex model, but this bias is small. Moreover, for the rejected cases, we note that the smallest value of the Wilks' statistic is ~$1.8 \times 10^{-7}$. Compared to the value of Wilks' statistic obtained in the case of the real catalog in favor of gv-ETAS5 model over gv-ETAS1 models (~$1.9 \times 10^{-283}$), the minimum value of the Wilks' statistic obtained for the synthetic case is indeed enormously larger. So, we conclude that the log-likelihood ratio test conducted in the previous section would not disproportionately favor the more complex gv-ETAS5 model over the gv-ETAS1 model if the latter were the underlying generating process. Finally, it is interesting to note that the estimated $\delta$ values when the gv-ETAS5 model was calibrated on the synthetic catalogs, and for which the Wilks' statistics favored the gv-ETAS5 model over the gv-ETAS1 model, range from 0 to 0.16 and cannot possibly explain the much larger delta value of 0.74 obtained for the Californian catalog.

## 4.4 Rationalization of previous observations of mainshock magnitude dependence of the exponent of the GR law

As mentioned in Section 1, the exponent $\beta$ of the GR law of the magnitude distribution of the direct aftershocks conditioned on the mainshock magnitude (and retrieved from the branching structure) has been observed to decrease with the mainshock magnitude [Zhuang et al., 2004], while a synthetic catalog that is simulated using the ETAS model (Equations 1 and 2) cannot reproduce this observation. Those findings can be rationalized using our (generalized Vere-Jones ETAS) Model 5: intuitively, the larger the magnitude ($M$) of the mainshock, the smaller the range of magnitude $m>M$ over which the second branch with $\beta_a + \delta$ exponent holds, and the more the estimation of the GR law is dominated by the first branch with $\beta_a - \delta$ exponent. As a result, we should expect to see a decrease of the overall apparent $\beta$ value of the direct aftershocks with increasing mainshock magnitude.

In the following, we proceed to quantitatively reconcile the observations of [Zhuang et al., 2004] using the generalised Vere-Jones ETAS model using the following strategy:

1. We first generate a synthetic catalogue using the gV-ETAS5 model using the parameters listed in Table 1.

2. We calibrate a standard ETAS (gV-ETAS1) model on it.

3. Having calibrated the standard ETAS model, we obtain the $P_{ij}$ values. We then define $N = \left\lceil \frac{M_{max} - M_1}{\delta M} \right\rceil$ non-overlapping mainshock magnitude bins of size $\delta M = 0.5$: $[M_1, M_1 + \delta M]$; $[M_1 + \delta M, M_1 + 2 \times \delta M]$…; $[M_1 + (N-1) \times \delta M, M_1 + N \times \delta M]$. Note that $M_1 = 3$ and $M_{max} = 7.3$ are the minimum and maximum magnitudes of the earthquakes present in the real catalog. We then group the earthquakes of the earthquake catalog in the aforementioned magnitude bins, and estimate the apparent $\beta$-value ($\beta_k$) of the $k^{th}$ mainshock magnitude bin according to the following formula:

$$\beta_k = \frac{\sum_{ij} P_{ij}^k}{\sum_{ij} P_{ij}^k \left(M_{ij}^k - M_c\right)} \tag{7}$$

In the above equation, $M_c = 3$ is the magnitude of completeness of the catalog; $M_{ij}^k$ is the magnitude of the $j^{th}$ earthquake that has been triggered by the $i^{th}$ earthquake present in the $k^{th}$ magnitude bin with a probability $P_{ij}^k$.

4. We repeat steps 1-3 a total of 100 times.

To be consistent with [Zhuang et al., 2004], we also calibrate the ETAS model (Equations 1 and 2) on the real Californian catalogue and, using the obtained triggering probabilities, compute the $\beta$-values conditioned on the mainshock magnitude using the methodology defined above.

Figure 3 displays the median estimates of the $b$ value $\left( = \frac{\beta}{\log(10)} \right)$ and its 95% confidence interval as a function of the mainshock magnitude (solid and dashed blue lines, respectively), obtained from synthetic catalogues generated using Model 5 (using the parameters listed in Table 1), their branching structure being inverted and approximated by the standard ETAS model. The red circles in the figure show similar estimates (i.e. after inversion with ETAS and using the associated branching structure) to the *b* values conditioned on the mainshock magnitude obtained from the real Californian catalogue. The horizontal solid black line in the figure shows the constant *b* value predicted by the ETAS model when applied to California. We also show the *b* values previously reported [Zhuang et al., 2004] for the JMA catalogue using black crosses. The dependence of the *b* value as a function of mainshock magnitude obtained from the real Californian catalog agrees very well with the trend that is empirically predicted by Model 5. A similar dependence can be observed in the empirical observations of [Zhuang et al., 2004], while the difference can be associated with the differences between the Californian and Japanese catalogues. The overall similarity of the magnitude dependence of the *b* values found in the ANSS and JMA catalogues points towards the universality of Model 5.

## 4.5 Could the known biases in the estimates of ETAS parameters spuriously lead to mainshock magnitude dependent b-values?

The parameter space of a spatial ETAS model is on a hyperplane of about two dimensions less than that of the nominal number of model parameters, hence there is considerable correlation between some of the parameters [Harte, 2016, Figs 10,11, Tables 1-3]. Hence, if one ignores some of the important characteristics of the underlying generating model, say a spatially variable background rate, the parameter correlation will forcibly introduce some spurious affects in other parameters that may be thought unrelated. Addition of extra structure into the model effectively constrains the model equations, and the model fitting is forced to use other parameters to account for various empirical characteristics, flexibility that it previously had because of the sloppy structural specification (discussed in Harte 2013, Appendix B). Thus, it is of paramount importance that we analyse if some of the known sources of biases in the estimates of the ETAS parameters can possibly explain the origin of magnitude dependent b-values of aftershocks as observed by us and other studies [Nichols and Schoenberg, 2014; Zhuang et al., 2004; Spassiani and Sebastiani, 2016]. There are several known sources of bias in the estimates of the ETAS parameters. Some of the important ones include:

1. Short-term aftershock incompleteness: several researchers [Hainzl, 2016; Kagan, 2004; Helmstetter et al., 2006; Peng et al., 2007] have argued that, at short times following large earthquakes, the base magnitude of completeness temporarily rises. As a result, earthquakes, which should have been recorded normally, might be missing in the short times following large earthquakes. Ignoring the short-term aftershock incompleteness leads to biases in the estimates of the ETAS parameters, as has been extensively demonstrated in the literature.

2. Finite faults: The distribution of aftershocks around a mainshock (epicenter or hypocenter) is far from isotropic. Instead, the aftershocks are distributed on and around the finite portion of faults that the mainshock has activated. However, in most formulations of the ETAS model (including our own), the mainshocks are treated as point sources, and aftershocks are distributed isotropically around them. Several works [Hainzl et al., 2008; Bach and Hainzl, 2012] have demonstrated that such simplifications introduce severe biases in the estimates of the ETAS parameters.

3. Spatial variation in parameters of the ETAS model: ETAS parameters have been known to show significant spatial variations in the current study region [Nandan et al., 2017].

The global estimates of the parameters cannot alone account for this spatial variation and are likely to be biased.

Although the list above may not be exhaustive, it nonetheless includes the three most important sources of biases in the estimates of the ETAS parameters. In our analysis so far, we have not accounted for any of the three above mentioned biases, and tests of appropriate null hypotheses could reveal if our observations are the results of the distortions that arise in the branching structure due to unaccounted biases.

### 4.5.1 Short-term aftershock incompleteness (STAI) simulations

In this test, we investigate if the biased branching structure, which results from unaccounted short-term aftershock incompleteness (STAI) during ETAS model calibration on a synthetic catalogue, could lead to the observation of mainshock magnitude dependent $\beta$-value of the direct aftershocks, as has been shown for the real catalogue. In this test, we simulate 1000 synthetic catalogues with known ETAS parameters and an assumed magnitude of completeness ($M_c$). In the simulated catalogues, we impose STAI to obtain observed incomplete catalogues. We treat these observed catalogues as if they were complete above $M_c$ to mimic real catalogues. Then we calibrate the global ETAS model on these observed catalogues and obtain the estimate of mainshock magnitude dependent $\beta$-value of direct aftershocks as proposed in Section 4.3.

To perform the ETAS simulations, one would ideally want to use realistic parameters. While it is tempting to assume that the parameters obtained upon the calibration of the ETAS model (values reported for the gV-ETAS1 in Table 1) on the real catalogue are realistic, it is important to remember that those parameters might have already been biased, as we have not considered the STAI during the calibration process. We thus have no strong priors for the estimates of realistic ETAS parameters. Nevertheless, using physical considerations, it might be possible to arrive at such an estimate, not considered to be the absolute truth, but good enough for our simulation purposes. While STAI should influence the estimates of all the parameters of the ETAS model, the two main parameters that should be affected the most are the exponent of the productivity law ($a$) and the exponent of the Omori law ($\omega$). It is reasonable to expect that both of these parameters should be underestimated as a result of unaccounted STAI. In the calibration of the ETAS model on the original catalogue, we have found that the estimates of $a$ and $\omega$ are 1.01 and 0.17, respectively. Assuming that these values could be underestimated, we increase the value of both these parameters to 2.38 ($a = \beta_a = 2.38$) and 0.3 respectively. Furthermore, we assume that all the other parameters of the ETAS model are respectively the same as found for the real catalogue: branching ratio or $n = 0.99$, $\beta_a = \beta_b = 2.38$, $c = 0.017 \; days$, $\rho = 0.53$, $d = 0.24 \; km^2$ and $\gamma = 1.11$. The only exception is the parameter $\mu$, which controls the rate of background earthquakes in the catalogue. We noticed that, if we use the original value of the parameter $\mu$ found in the real catalogue, we do not obtain as many earthquakes as has been observed in the actual catalogue, in the same time duration and spatial extent. This mismatch between the simulated number of earthquakes and the observed number of earthquakes in the real catalogue does not contradict the quality of the initiation calibration, which reproduces correctly the number statistics of the real catalogue, as a result of a suitable interplay between the fitted parameters. By modifying the parameters $a$ and $\omega$ only, this interplay is disrupted. Further, if we impose additionally the STAI, the number of earthquakes in the simulated catalogue would also decrease. To ensure that the number of earthquakes in the simulated catalogue after imposing the STAI is on average in agreement with the number of earthquakes in the original catalogue, we consider $\mu$ as a free parameter and optimise it to maximise the average agreement between the simulated and the real total number of earthquakes. We find that, by setting $\mu$ to nearly four times its original value in the real catalogue, the numbers of earthquakes in the simulated and real catalog match on average. With the above considerations, we proceed with the simulations as follows:

1. We simulate $N$ background earthquakes, where is $N$ is a random number simulated from a Poisson distribution, with expected value being $\mu TA$. T and A are the time duration and area of the region in which the real earthquakes are distributed. The times of the background earthquakes are simulated uniformly randomly between $[0, T]$, and the background earthquakes are also distributed uniformly randomly in space. We simulate the magnitude of the background earthquakes using the Gutenberg Richter (GR) law with an exponent, $\beta_b = 2.38$, $M_c=3$ and maximum magnitude ($M_{max}$) being 8.5. All the background earthquakes, $(t_i, x_i, y_i, M_i)$ are treated as the parent earthquakes for the next generation of aftershocks.

2. For the $i^{th}$ parent earthquake, the number of direct aftershocks ($M \geq 3$) is given by $N_i$, where $N_i$ is a random number simulated from a Poisson distribution with expected value equal to $Ke^{a(M_i-3)}$. Note that the productivity coefficient $K$ can be easily obtained, given the value of the minimum and the maximum magnitudes of earthquakes that can occur (3 and 8.5 respectively), the branching ratio ($n = 0.99$), and the exponents of the productivity and the GR law ($a = \beta_a = 2.38$).

3. The times and locations of the $N_i$ aftershocks are simulated from the Omori kernel and spatial kernel whose parameters are $\omega$ and $c$ (0.3 and 0.017 days respectively), and $\rho$, $d$ and $\gamma$ ($0.53, 0.24\ km^2$ and 1.11 respectively).

4. The magnitudes of the $N_i$ aftershocks are simulated using the GR law, $\beta_a$ $M_c$ and $M_{max}$ being 2.38, 3 and 8.5 respectively. Notice that no magnitude-dependent effect on $\beta_a$ is introduced here.

5. This simulation process of direct aftershocks (steps 2 to 4) is repeated for all the parent earthquakes.

6. The combined list of all the direct aftershocks simulated in Step 5 is treated as the new list of parent earthquakes, and the old list is discarded.

7. We repeat steps 2 to 6 until no more earthquakes are available in the parent list.

8. We then impose the STAI (using the equation $M_c(t, M) = M - 4.5 - 0.75 \log_{10} t$ proposed by Helmstetter et al. [2006]) on the simulated catalogue to obtain the observed catalogue of earthquakes. In this formulation of STAI, $M_c(t, M)$ is the magnitude of completeness at the time $(t)$ after an earthquake of magnitude $M$. Thus, at each time, each past event induces a different value of $M_c$. At that time, the completeness magnitude of the catalogue is thus the maximum of those $M_c's$. At each time, all events below the current max($M_c$) value are thus removed from the observed catalog, despite the fact that they do participate in the triggering of the following events.

9. We simulate 1000 such catalogues.

We find that, with the parameter settings assumed for the STAI simulations, in 95% of the simulated catalogues, 43% to 93% of the earthquakes are missing and half of the simulations miss more than 73% earthquakes. Although such degree of incompleteness might not be realistic, it nevertheless serves as an extreme null hypothesis to test if the mainshock magnitude dependent b-value of direct aftershocks might have a spurious origin or not.

**4.5.2 Hybrid simulations combining Finite Fault (FF) and STAI**

In this test, we want to investigate if the biased branching structure resulting from the unaccounted finiteness and anisotropy of the faults as well as the STAI during ETAS model calibration on synthetic catalogues could lead to the spurious occurrence of the mainshock magnitude dependent β-value of the direct aftershocks. The synthetic catalogues for this test are generated in the same way and with the same parameters as in Section 4.4.1, except for one main difference. During simulations, the earthquakes are not assumed to be point sources as in Section 4.4.1. Each earthquake is treated as a finite source whose length is obtained from Wells and Coppersmith equations [Wells and Coppersmith, 1994] for purely strike-slip earthquakes ($L = 10^{0.74M-3.55}$). Furthermore, we treat each earthquake as a line source with the earthquake's epicentre being the mid-point of the line segment. The line segment is assumed to be randomly oriented in space and has no memory of the orientation of the parent earthquake that triggered it. For a background earthquake, the orientation of the line segment is in accordance with the orientation of the far field tectonic loading (i.e. along a preassigned, constant direction).

One might argue that, even in 2D, the fault cannot be represented by a line unless its dip is equal to 90°. However, two reasons are justifying our simplification. First, assuming the projection of the fault plane to be a line segment in 2D leads to the most extreme anisotropy. Furthermore, assuming the projection to be a line makes the simulation task much easier, which would have to be otherwise performed in 3D. To distribute $N = Ke^{a(M-3)}$ aftershocks around a source earthquake with a given fault line, we first discretise the fault line with $N_p$ points, which we refer to as 'subsources'. The number of subsources depends on the magnitude of the source earthquake being discretised and also on the minimum magnitude (= 3) of the earthquake according to $N_p = \lfloor 10^{0.74(M-3)} \rfloor$. Each of the subsources along a fault segment are treated as isotropic and virtual point sources, each of them generating $\frac{N}{N_p}$ triggered earthquakes. The distances of these $\frac{N}{N_p}$ earthquakes to their mother subsource are simulated from the spatial kernel of the isotropic ETAS model with the same parameters as in the case of STAI simulations. However, the parent magnitude value in the original formulation of the kernel is replaced by the magnitude of the original source event. New fault segments are then introduced at the locations of the first generation aftershocks, and the process is repeated for the next generations.

In Figure S1, we show the spatial distribution of earthquakes generated in an illustrative simulation. Insets 1 and 2 in panel A of the figure illustrate clearly that the spatial distribution of earthquakes in regions 1 and 2 are far from isotropic, as would have been the case had the aftershocks been simulated without accounting for the finite size and anisotropy of events.

### 4.5.3 Hybrid simulations combining spatially variable ETAS (SV-ETAS), Finite Faults and STAI

In this test, we investigate the combined influence of unaccounted spatial variations in the ETAS model parameters, finite faults and short-term aftershock incompleteness during the calibration of the ETAS model on the synthetic catalogues. For these simulations, we use the spatially variable ensemble estimates of ETAS model parameters $\left(\mu, n, c, \omega, \alpha = \frac{a}{\log(10)}, b-value = \frac{\beta}{\log(10)}\right)$, which our current inversion algorithm is designed to invert (see Nandan et al. [2017] for details on the original method). The spatially variable estimates of the parameters are shown in Figure S2**.** In our current implementation of the inversion method, we assume that the parameters of the spatial kernel, $(d, \rho, \gamma)$, are global and their estimated values are respectively 0.20, 0.58 and 1.21. To simulate the catalogues, we use the following algorithm:

***1.*** We first simulate the background earthquakes. To do this, we first draw a number $U_i$ uniformly at random from the interval [0,1]. We then compare it to the independence probability $IP_i$ of the $i^{th}$ earthquake in the catalogue. Note that, $IP_i$ quantifies the likelihood that the $i^{th}$ earthquake in the catalog is a background earthquake and is obtained as an output of the calibration of the ETAS model on the real catalog [Nandan et al., 2017]. If $U_i \leq IP_i$, then the location of the $i^{th}$ earthquake is selected as the location of a background event in the simulation. We repeat this process for all the earthquakes in the catalog. For the selected seed events, we discard their actual time and magnitude information and only retain their location information. To these background earthquakes, we assign times by drawing random numbers uniformly from the interval [0,$T$]. We then simulate the magnitude of the $i^{th}$ background earthquakes using the following equation:

$$m_i = \frac{-log(1 - U_i^1)}{\beta_i} + M_c \qquad (8)$$

In the above equation, $U_i^1$ is a number drawn uniformly at random from the interval [0,1]; $\beta_i$ is the ensemble estimate of the exponent of the GR law at the location the background earthquake and $M_c$ is the magnitude of completeness, which is set to 3 for our simulations. All background earthquakes are treated as parent earthquakes for the next generation.

***2.*** An earthquake with a magnitude $m_i$ can trigger on average $K_i \, e^{[a_i(m_i - M_c)]}$ aftershocks with magnitude larger than $M_c = 3$. $K_i$ and $a_i$ represent the ensemble estimate of the productivity parameters at the location of the $i^{th}$ earthquake.

***3.*** The times of those aftershocks can be simulated according to the following equation:

$$t_{ji} = c_i \left\{ \left(1 - U_j^2\right)^{-\frac{1}{\omega_i}} - 1 \right\} + t_i \qquad (9)$$

In equation (9), $t_{ji}$ is the time of the $j^{th}$ aftershock that has been triggered by the $i^{th}$

earthquake in the training catalogue, which is counted from the time of the $i^{th}$ earthquake; $U_j^2$ is a random number drawn uniformly in the interval [0,1]; $c_i$ and $\omega_i$ are the ensemble estimate of the parameters of the Omori kernel at the location of the $i^{th}$ earthquake that occurred at time $t_i$.

*4.* To simulate the location of aftershocks around the $i^{th}$ earthquake, we use the simulation strategy developed for finite fault simulations.

*5.* We then simulate the magnitude of the direct aftershocks using the same strategy we have used for background earthquakes in Step 1.

*6.* This process of simulation of direct aftershocks (steps 2 to 4) is repeated for all the parent earthquakes.

*7.* The combined list of all the direct aftershocks simulated in Step 6 is treated as the new list of parent earthquakes, and the old list is discarded.

*8.* We repeat steps 2 to 7 until no more earthquakes are available in the parent list.

*9.* We use the Step 8 of the algorithm proposed in Section 4.4.1 to remove the earthquakes that fall below the time-dependent magnitude of completeness determined by the Helmstetter's incompleteness equation.

*10.* We repeat this process 1000 times to obtain 1000 synthetic catalogues.

### 4.5.4 Results of the three simulations

Upon calibration of the global ETAS model (gV-ETAS1), which assumes that the simulated catalogues are complete and the aftershocks are distributed isotropically around the mainshocks, we obtain the triggering probability matrix for all the simulated catalogues in the three simulation types. Using these triggering probability matrices and the method proposed in Section 4.3, we then obtain the mainshock magnitude dependent b-value of direct aftershocks. In Figure 3, we show the 50, 2.5 and 97.5 quantiles of the estimates of the b-values obtained for different mainshock magnitude bins for the three simulation types using error bars. We find that despite all the biases in the estimates of the parameters that would have occurred in the three simulation types, in none of them we observe the same dependence as has been found in the real catalogues by Zhuang et al. [2004] and us. Among the tested models, only gV-ETAS5 can predict this dependence.

## 5 Discussion and Conclusions

### 5.1 On-fault versus off-fault aftershocks.

One could argue that a possible origin of the kinked GR law is the existence of two types of aftershocks, those on-fault and those off-fault [Sornette et al., 1990]. Here, we consider as on-fault aftershocks those nucleating on the same rupture plane as their ancestor. Assuming that it is unlikely for the on-fault earthquake sizes to exceed the size of the trigger, they would thus contribute more to the $\beta-\delta$ branch for $m < m_i$ of the kinked GR law. Off-fault aftershocks, which would not suffer a priori such a limiting size, would thus contribute more to the $\beta+\delta$ branch for $m > m_i$. To test if this idea has any explanatory power, using empirical relationships between magnitude $M$ and rupture size $L$ [Wells and Coppersmith, 1994], we separated triggered events into two different categories: those located at distances to their trigger $< L(M)$, where $M$ is the magnitude of the trigger and $L(M)$ is the length of its rupture, and those located at distances $>$ $NxL(M)$, where $N$=2, 3,...,20. Aftershocks located at distances $> L(M)$ and $< NxL(M)$ were discarded to limit the possibilities of mixing on- and off-fault events. Using the branching structure and the same stacked representation as in Figure 1, we found, for all $N$, that the same kinked distribution holds for both subsets of triggered events, even at very large distances from the trigger. This result clearly excludes the on-fault and off-fault aftershocks interpretation and suggests that the explanation of the kink should not be sought in the local mechanical perturbations due to the triggering event, but rather stems from a nonlocal, global criterion or conservation law at the scale of the full system.

On a side note, this result also confirms that short-term incompleteness issues do not pollute our analyses: such incompleteness is clearly a space-time sampling process, as the space-time density of events is larger close to the main event hypocenter than away from it. In the case of large events, long-distance triggered events will also be recorded by a different set of stations than small-distance ones. It thus follows that STAI problems should, in general, be more acute close to the main events. As we observe the same signature far away from main events and close to them, we can conclude that the kinked magnitude distribution is not due to such a sampling and detection problem, and that the magnitude of the triggering events does not play any role in this phenomenology.

## 5.2 Symmetry principles, critical phenomena, and self-organised criticality.

The theorem of Pierre Curie [1894] enunciates that 'the symmetries of the causes are to be found in the effects'. For instance, a sample of homogeneous material subjected to a triaxial state of stress will display displacement fields that have the same symmetry as the stress tensor applied at the external boundaries of the specimen, at least below its failure threshold. It is anyway widely observed that systems composed of numerous interacting elements will spontaneously break their symmetry at the macroscopic level when the externally applied control parameter passes over a threshold. The order parameter (a macroscopic property measured on the system) then undergoes a bifurcation, for instance, in the case for Heisenberg spin glasses [Stanley, 1971] in the absence of an external magnetic field (so that no preferential direction is imposed). At temperatures above Curie's critical value $T_c$, thermal fluctuations constantly overcome the coupling between neighbouring spins, so that the macroscopic magnetisation vector is zero: all spins are randomly oriented so that the global symmetry is conserved. When decreasing the temperature, thermal fluctuations become weaker and, below $T_c$, a finite magnetisation appears: spins now tend to be oriented in a preferred direction due to their coupling, thus spontaneously breaking the isotropic symmetry. Right at the critical point $T_c$, the system is scale-invariant, displaying many power-law distributions and correlations, but still with a zero net magnetisation. If the temperature is then slightly decreased below $T_c$, Goldstone modes appear [Goldstone et al., 1962; Lange, 1966], consisting in very long wavelength waves of spin oscillations, whose effect is to destroy the long-range order in the system at minimal energy cost (theoretically null, i.e. gapless, in Heisenberg's model). In other words, those modes attempt to restore the lost symmetry, as long as the system is not too far from the critical point. Goldstone modes are usually observed at the scale of quantum phenomena, but similar observations of broken symmetries and their partial restoration via Goldstone modes also hold at larger scales in soil and granular matter mechanics, for instance. When a cylinder of such material is subjected to an axisymmetric state of stress, a bifurcation occurs when one of the conjugate slip surfaces finally localises all the deformation after the load has reached a critical value, to finally lead to the failure of the sample along that single plane. Right at the bifurcation, surprisingly symmetric patterns of failure surfaces may emerge: some developing as revolution cones around the main stress axis [Desrues et al., 1996], or as spiral staircases gyrating around that same axis [Desrues et al., 1991; Evesque and Sornette, 1993]. Both kinds of structures attempt to preserve, at least partly, the initial cylindrical symmetry although it will be ultimately destroyed when loading further the sample, beyond the critical point.

For several decades, earthquakes and faulting have been considered as shining examples of self-organised criticality (see, e.g. [Sornette, 1991; 2006]). In the latter class of processes, under conditions of slow loading (as is the case for plate tectonics), a system featuring many interacting elements will spontaneously reach a critical state, without the need to finely tune any external parameter, and will remain in that state, a stable and attractive point of the collective dynamics. This broad concept allows one to rationalise the numerous power-laws observed in earthquakes phenomenology that are used into the standard ETAS kernels (Equation (2)).

Pushing the concept further, we may consider an initially intact macroscopic tectonic domain. Progressive loading at its boundaries first induces a homogeneous coarse-grained deformation, which can be decomposed into the combination of two tensorial order parameters: (i) a symmetric second order strain tensor, whose orthorhombic symmetry class is the same as the one of the control parameter, the applied stress tensor, in agreement with Curie's principle; and (ii) a null but anti-symmetric rigid rotation tensor, which can equivalently be replaced by a rotation vector, i.e belonging to a completely different monoclinic symmetry class. When the first faults begin to nucleate, they generally develop in conjugate systems, symmetric with respect to the maximum stress axis [Anderson, 1905]. Sliding over any fault of a given system increments both the strain and rigid rotation tensors. The associated rotation vector is collinear to $\vec{\varphi} = \vec{u} \times \vec{n}$, where $\vec{u}$ and $\vec{n}$ are respectively the slip vector and the unit vector normal to the fault. Sliding over any fault of the conjugate system also increments the strain tensor but induces a rigid rotation of opposite sign.

Fault networks grow in a self-organised manner due to the accumulation of earthquakes with time and their mutual interactions, until the critical state is reached [Sornette et al., 1990; Sornette, 1991], which is a fixed point. In this view, the static effect of a background earthquake, which is nothing but a finite slip increment along a finite plane due to the remote loading, can also be split into separate contributions to the strain and rigid rotation tensors, the amplitude of both being proportional to the scalar seismic moment of the event [Kostrov, 1974; Mukhamediev and Brady, 2002; Legrand, 2003]. The rigid rotation is the order parameter quantifying the departure from orthorhombic symmetry of the displacement field (as well as measuring the amount of strain localisation). Given that the system stands exactly at the critical point when a fault nucleates, a background event can thus be viewed as an embryonic spontaneous break of symmetry. A Goldstone-like mode is thus expected to emerge to counteract this symmetry loss, i.e. to induce a counter-rotation of similar amplitude. The most natural way to achieve the latter is by triggering another event with, preferably, a similar magnitude on another fault with conjugate displacement. This is, we propose, how the correlation of magnitudes appears. Yet, that Goldstone-like event has also to account for the presence of noise and the proper symmetries of rupture propagation, the latter being largely controlled by the symmetry of the pre-existing fault network and the equations of elastodynamics. Their scale invariance certainly contributes to the fact that the final size of isolated events will usually obey a pure GR law. In the case of a triggered event, this distribution will thus be modulated by the Goldstone-like mode which, we suggest, explains the kink observed in the GR distribution of direct aftershocks. As each of the directly (and identically independently distributed) triggered events attempts independently to counteract the rotation due to the triggering parent, some of them will also trigger their own aftershocks, and so on, in a cascade of Goldstone-like modes.

The proposed Goldstone-based mechanism may also rationalise other surprising observations. In particular, most observations of direct triggering report very large distances between the triggering and triggered events, where the transferred stress (and thus strain energy) is vanishingly small. Many triggered events even occur within zones where the static Coulomb stress is predicted to be at a lower level than before the mainshock occurred [Nandan et al., 2016]. Moreover, both usual static and dynamic stress triggering hypotheses alone do not provide any explanation for the observed magnitude correlations, while the Goldstone-like mode cascade may rationalise these surprising observations.

This novel formulation of seismicity, stemming from the analysis of the symmetries of the control parameter (here, the externally applied tensorial stress field) thus suggests the existence of significant differences in the origin of mainshocks and direct aftershocks: the former occur primarily to accommodate the symmetric stress tensor at the boundaries of the system, and are distributed according to a standard GR law; the latter occur to cancel the rigid rotation induced by the former and are distributed according to a different, kinked GR law. It should anyway be stressed that this dynamical behaviour also requires the system to reach (and stay close to) a critical point.

A full test of the theory would be possible if we knew the failure plane for each event in an earthquake catalogue so that individual rotation vectors could be computed. Unfortunately, focal mechanisms do not provide such information due to the ambiguity associated with the symmetric double-couple representation. Furthermore, seismicity-based fault reconstructions, despite major improvements in the last decade [Ouillon and Sornette, 2011; Wang et al., 2013; Kamer, 2015], are not sufficiently advanced to provide the necessary data, and would certainly require more objectively located catalogues of events [Kamer et al., 2017] and focal mechanisms. Nevertheless, we notice that conjugate patterns of faulting have already been observed on a very wide range of space and time scales. In the spatial domain, such patterns exist from large-scale faulting (i.e. the long-time scale as well) down to laboratory samples [Fossen, 2016]. On intermediate time scales, they are also clearly observable on structures delineated by earthquake catalogues, such as in the San Jacinto area [Ross et al., 2017], for instance. At the scale of aftershock sequences, one of the most famous examples is one of the Elmore Ranch (M=6.2) and Superstition Hills (M=6.6) earthquake sequences [Hudnut et al., 1989], with two conjugate events of similar magnitudes occurring on November 24, 1987. At even smaller time scales, Meng et al. [2012] showed that the 2012, $M_w$=8.6 event in Sumatra had a very complex rupture process involving several conjugate fault segments. Fukuyama [2015] reports several other examples of conjugate ruptures occurring during the same event or within short-term aftershock sequences (see also Hauksson et al. [2002] for the Hector Mine event). Those observations at multiple scales are indeed consistent with the underlying idea of scale invariance of the faulting process and call for a complete re-examination of symmetry patterns in faulting. Beyond the peculiar problem of earthquakes physics, we would like to point out that nonlinear interactions of Goldstone modes have also been invoked as one of the possible routes to self-organised criticality [Obukhov, 1990] among many other mechanisms [Sornette, 2006]. However, the predicted size distributions of associated events is then a pure GR law. The work presented in the present article may thus open new doors and fascinating ways to explore self-organised critical systems featuring control and order parameters belonging to different and conflicting symmetry classes.

### 5.3 Implications for earthquakes forecasting and prediction.

Our main result is that the distribution of earthquake magnitudes in empirical catalogues exhibits three distinct $\beta$-values: $\beta_b$, $\beta_a - \delta$ and $\beta_a + \delta$, instead of a single one, and that the two last values express a correlation between the magnitudes of triggered events with that of the triggering earthquake. How can this be reconciled with the fact that the overall magnitude distribution seems well-explained by a single exponent $\beta$? Saichev and Sornette [2005] have shown that, notwithstanding the existence of three distinct $\beta$-values in the generalized Vere-Jones ETAS (Equations (3)-(4)), the whole set of aftershocks of all generations following a given initiating event is distributed according to a pure GR law with a single exponent $\beta_a$. An important message is therefore that the existence of several $\beta$-values and underlying correlations between magnitudes do not become visible without a rigorous separation of each seismicity layer (background events, 1st generation aftershocks, 2nd generation aftershocks, and so on), as done in the present work.

These correlations between magnitudes can be expected to have a significant impact on the performance of statistically-based forecasting algorithms mentioned above [Zechar et al., 2013; Rhoades et al., 2014]. Up to now, these correlations have either been ignored or have been incorporated in indirect ways in forecasting models such as in Ogata et al. [2018], so that they only lead to small probability gain over models that ignore pairwise magnitude correlations. Other studies such as Lippiello et al. [2012] have also pointed towards similar outcomes as found by Ogata et al. [2018] as they noticed that pairwise magnitude correlations almost vanish after a few tens of minutes. Only rigorously conducted pseudo prospective forecasting experiments, which we plan to conduct in future studies, could reveal if the magnitude correlations such as the one inherent in the gv-ETAS5 model would lead to significant forecasting gains over the classical ETAS model.

Our results also call for a re-examination of previous claims that the *b*-value of the GR law (which is given by $\frac{\beta}{\log 10}$) can be used as a tool to predict upcoming large earthquakes [Scholz, 2015] or to monitor space-time variations of the stress level [Schorlemmer et al., 2005; Kamer and Hiemer, 2015]. Our findings show that *b*-values obtained without declustering may not have the same physical meaning. It is likely that correlations between *b*-values and stress or strain regimes are more meaningful, if they exist, when using the exponent $\beta_b$ associated with the background events.

## Acknowledgement

The dataset used in this study can be obtained from the website.

http://www.quake.geo.berkeley.edu/anss/catalog-search.html.

S.N. wishes to thank Y. Ben-Zion (editor), the anonymous associate editor and three anonymous reviewers for critical suggestions and discussions, which led to significant improvement of the manuscript.

## References

1. Aki, K. (1965). Maximum likelihood estimate of b in the formula log N= a-bM and its confidence limits. *Bull. Earthq. Res. Inst., Tokyo Univ.*, *43*, 237-239.
2. Anderson, E. M. (1905). The dynamics of faulting. *Transactions of the Edinburgh Geological Society*, *8*(3), 387-402.
3. Bach, C., & Hainzl, S. (2012). Improving empirical aftershock modeling based on additional source information. *Journal of Geophysical Research: Solid Earth*, *117*(B4).
4. Busemeyer, J. R., & Wang, Y. M. (2000). Model comparisons and model selections based on generalization criterion methodology. *Journal of Mathematical Psychology*, *44*(1), 171-189.
5. Chernoff, H. (1954). On the distribution of the likelihood ratio. *The Annals of Mathematical Statistics*, 573-578.
6. Cudeck, R., & Browne, M. W. (1983). Cross-validation of covariance structures. *Multivariate Behavioral Research*, *18*(2), 147-167.
7. Curie, P. (1894). Sur la symétrie dans les phénomènes physiques, symétrie d'un champ électrique et d'un champ magnétique. *Journal de physique théorique et appliquée*, *3*(1), 393-415.
8. Desrues, J., Chambon, R., Mokni, M. & Mazerolle, F. (1996): Void ratio evolution inside shear bands in triaxial sand specimens studied by computed tomography, Géotechnique, 46, No. 3, 529-546.
9. Desrues J., Mokni M. And Mazerolle F. (1991): Tomodensitométrie et la localisation sur les sables. 10th E.C.S.M.F.E., Florence, Balkema eds, pp. 61-64-1991.
10. Evesque, P. and D.Sornette, A dynamical system theory of large deformations and patterns in non-cohesive solids, Phys.Lett. A 173, 305-310 (1993).
11. Fossen, H. (2016). *Structural geology*. Cambridge University Press.

12. Fukuyama, E. (2015), Dynamic faulting on a conjugate fault system detected by near-fault tilt measurements, *Earth Planets Space*, *67*(1), 1–10, doi:10.1186/s40623-015-0207-1.

13. Goldstone, J., Salam, A., & Weinberg, S. (1962). Broken symmetries. *Physical Review*, *127*(3), 965.

14. Gutenberg, B., & Richter, C. F. (1944). Frequency of earthquakes in California. *Bulletin of the Seismological Society of America*, *34*(4), 185-188.

15. Hainzl, S., Christophersen, A., & Enescu, B. (2008). Impact of earthquake rupture extensions on parameter estimations of point-process models. *Bulletin of the Seismological Society of America*, *98*(4), 2066-2072.

16. Hainzl, S. (2016), Apparent triggering function of aftershocks resulting from rate-dependent incompleteness of earthquake catalogs, *Journal of Geophysical Research: Solid Earth*, (9), doi:10.1002/2016jb013319.

17. Hainzl, S., O. Zakharova, and D. Marsan (2013), Impact of Aseismic Transients on the Estimation of Aftershock Productivity ParametersImpact of Aseismic Transients on the Estimation of Aftershock Productivity Parameters, *B Seismol Soc Am*, *103*(3), 1723–1732, doi:10.1785/0120120247.

18. Hainzl, S. (2016), Rate-Dependent Incompleteness of Earthquake Catalogs,*Seismol Res Lett*, *87*(2A), 337–344, doi:10.1785/0220150211.

19. D. S. Harte; Bias in fitting the ETAS model: a case study based on New Zealand seismicity, *Geophysical Journal International*, Volume 192, Issue 1, 1 January 2013, Pages 390–412, https://doi.org/10.1093/gji/ggs026

20. Harte (2016), Model parameter estimation bias induced by earthquake magnitude cut-off, *Geophys J Int*, *204*(2), 1266–1287, doi:10.1093/gji/ggv524.

21. Hauksson, E., L. Jones, and K. Hutton (2002), The 1999 Mw 7.1 Hector Mine, California, Earthquake Sequence: Complex Conjugate Strike-Slip Faulting, *B Seismol Soc Am*, *92*(4), 1154–1170, doi:10.1785/0120000920.

22. Helmstetter, A., & Sornette, D. (2002). Subcritical and supercritical regimes in epidemic models of earthquake aftershocks. *Journal of Geophysical Research: Solid Earth*, *107*(B10).

23. Helmstetter, A., Sornette, D., & Grasso, J. (2003). Mainshocks are aftershocks of conditional foreshocks: How do foreshock statistical properties emerge from aftershock laws. *Journal of Geophysical Research: Solid Earth (1978–2012)*. doi:10.1029/2002JB001991

24. Helmstetter, A., & Sornette, D. (2003). Predictability in the epidemic-type aftershock sequence model of interacting triggered seismicity. *Journal of Geophysical Research: Solid Earth*, *108*(B10).

25. Helmstetter, A., Y. Kagan, and D. Jackson (2006), Comparison of Short-Term and Time-Independent Earthquake Forecast Models for Southern California,*B Seismol Soc Am*, *96*(1), 90–106, doi:10.1785/0120050067.

26. Hudnut, Seeber, and Pacheco (1989), Cross-fault triggering in the November 1987 Superstition Hills Earthquake Sequence, southern California, *Geophys Res Lett*, *16*(2), 199–202, doi:10.1029/gl016i002p00199.

27. Kagan, Y. (2004), Short-term properties of earthquake catalogs and models of earthquake source, *Bulletin of the Seismological Society of America*, *94*(4), 1207–1228, doi:10.1785/012003098.

28. Kamer, Y., & Hiemer, S. (2015). Data-driven spatial b value estimation with applications to California seismicity: To b or not to b. *Journal of Geophysical Research: Solid Earth*, *120*(7), 5191-5214.

29. Kamer, Y., Kissling, E., Ouillon, G., & Sornette, D. (2017). KaKiOS16: A Probabilistic, Nonlinear, Absolute Location Catalog of the 1981–2011 Southern California Seismicity. *Bulletin of the Seismological Society of America*, *107*(5), 1994-2007.

30. Kamer, Y (2015), Magnitude frequency, spatial and temporal analysis of large seismicity catalogs: The Californian Experience, [online] Available from: https://www.research-collection.ethz.ch/bitstream/handle/20.500.11850/155463/eth-48708-01.pdf

31. Kostrov, V. V. (1974). Seismic moment and energy of earthquakes, and seismic flow of rock. *Izv. Acad. Sci. USSR Phys. Solid Earth*, *1*, 23-44.

32. Lange, R. V. (1966). Nonrelativistic theorem analogous to the Goldstone theorem. *Physical Review*, *146*(1), 301.

33. Legrand, D. (2003), A Short Note on the Selection of the Fault Plane, the Seismic Moment Tensor, the Strain and Rotational Tensors, and the Gradient of Displacement, *B Seismol Soc Am*, *93*(2), 946–947, doi:10.1785/0120020172.

34. Lippiello, E., Godano, C., & De Arcangelis, L. (2012). The earthquake magnitude is influenced by previous seismicity. *Geophysical Research Letters*, *39*(5).

35. Marsan, D., & Lengline, O. (2008). Extending Earthquakes' Reach Through Cascading. *Science*, *319*(5866), 1076-1079.

36. Meng, J.-P. Ampuero, Stock, Duputel, Luo, and Tsai (2012), Earthquake in a Maze: Compressional Rupture Branching During the 2012 Mw 8.6 Sumatra Earthquake, *Science*, *337*(6095), 724–726, doi:10.1126/science.1224030.

37. Mukhamediev, S. A., & Brady, B. H. G. (2002). On methods of the macro-stress determination by fault-slip inversions. *Structural integrity and fracture. Bulkema Publishers, Lisse, The Netherlands*, 277-281.

38. Nandan, S., Ouillon, G., Woessner, J., Sornette, D., & Wiemer, S. (2016). Systematic assessment of the static stress triggering hypothesis using interearthquake time statistics. *Journal of Geophysical Research: Solid Earth*, *121*(3), 1890-1909.

39. Nandan, S., Ouillon, G., Wiemer, S., & Sornette, D. (2017). Objective Estimation of Spatially Variable Parameters of Epidemic Type Aftershock Sequence Model: Application to California. *Journal of Geophysical Research: Solid Earth*.

40. Nandan, S (2017), Towards a Physics Based Epidemic Type Aftershock Sequence Model, [online] Available from: https://www.research-collection.ethz.ch/handle/20.500.11850/213570

41. Nichols, K., & Schoenberg, F. P. (2014). Assessing the dependency between the magnitudes of earthquakes and the magnitudes of their aftershocks. *Environmetrics*, *25*(3), 143-151.

42. Obukhov (1990), Self-organized criticality: Goldstone modes and their interactions., *Phys Rev Lett*, *65*(12), 1395–1398, doi:10.1103/physrevlett.65.1395.

43. Ogata, Y. (1988). Statistical models for earthquake occurrences and residual analysis for point processes. *Journal of the American Statistical association*, *83*(401), 9-27.

44. Ogata, Y. (1998), Space-Time Point-Process Models for Earthquake Occurrences, *Ann I Stat Math*, *50*(2), 379–402, doi:10.1023/A:1003403601725.

45. Ogata, Y., Katsura, K., Tsuruoka, H., & Hirata, N. (2018). Exploring magnitude forecasting of the next earthquake. *Seismological Research Letters*.

46. Omi, T., Y. Ogata, Y. Hirata, and K. Aihara (2014), Estimating the ETAS model from an early aftershock sequence, *Geophys Res Lett*, *41*(3), 850–857, doi:10.1002/2013GL058958.

47. Ouillon, G., & Sornette, D. (2011). Segmentation of fault networks determined from spatial clustering of earthquakes. *Journal of Geophysical Research: Solid Earth*, *116*(B2).

48. Peng, Z., J. Vidale, M. Ishii, and A. Helmstetter (2007), Seismicity rate immediately before and after main shock rupture from high-frequency waveforms in Japan, *Journal of Geophysical Research*, *112*(B3), doi:10.1029/2006JB004386.

49. Pinheiro, J., and D. Bates (2000), *Mixed-Effects Models in S and S-PLUS*, springer.

50. Reasenberg, P. (1985). Second-order moment of central California seismicity, 1969–1982. *Journal of Geophysical Research: Solid Earth*, *90*(B7), 5479-5495.

51. Rhoades, D. A., Gerstenberger, M. C., Christophersen, A., Zechar, J. D., Schorlemmer, D., Werner, M. J., & Jordan, T. H. (2014). Regional earthquake likelihood models II: Information gains of multiplicative hybrids. *Bulletin of the Seismological Society of America*, *104*(6), 3072-3083.

52. Ross, Z., E. Hauksson, and Y. Ben-Zion (2017), Abundant off-fault seismicity and orthogonal structures in the San Jacinto fault zone, *Sci Adv*,*3*(3), e1601946, doi:10.1126/sciadv.1601946.

53. Saichev, A., & Sornette, D. (2005). Vere-Jones' self-similar branching model. *Physical Review E*, *72*(5), 056122.

54. Schoenberg, F., A. Chu, and A. Veen (2010), On the relationship between lower magnitude thresholds and bias in epidemic-type aftershock sequence parameter estimates, *J Geophys Res Solid Earth 1978 2012*, *115*(B4), doi:10.1029/2009JB006387.

55. Scholz, C. H. (2015). On the stress dependence of the earthquake b value. *Geophysical Research Letters*, *42*(5), 1399-1402.

56. Schorlemmer, D., Wiemer, S., & Wyss, M. (2005). Variations in earthquake-size distribution across different stress regimes. *Nature*, *437*(7058), 539-542.

57. Schorlemmer, D., & Gerstenberger, M. C. (2007). RELM testing center. *Seismological Research Letters*, *78*(1), 30-36.

58. Seif, S., A. Mignan, J. Zechar, M. Werner, and S. Wiemer (2017), Estimating ETAS: The effects of truncation, missing data, and model assumptions, *J Geophys Res Solid Earth*, *122*(1), 449–469, doi:10.1002/2016jb012809.

59. Self, S. G., & Liang, K. Y. (1987). Asymptotic properties of maximum likelihood estimators and likelihood ratio tests under nonstandard conditions. *Journal of the American Statistical Association*, *82*(398), 605-610.

60. Sornette, D., Davy, P., & Sornette, A. (1990). Structuration of the lithosphere in plate tectonics as a self-organized critical phenomenon. *Journal of Geophysical Research: Solid Earth*, *95*(B11), 17353-17361.

61. Sornette, D. (1991). Self-organized criticality in plate tectonics. In *Spontaneous formation of space-time structures and criticality*, edited by T. Riste and D. Sherrington, Dordrecht, Boston, Kluwer Academic Press (1991), volume 349, p. 57-106.

62. Sornette, D., & Werner, M. J. (2005). Constraints on the size of the smallest triggering earthquake from the epidemictype aftershock sequence model, Båth's law, and observed aftershock sequences. *Journal of Geophysical Research: Solid Earth*, *110*(B8).

63. Sornette, D. (2006). *Critical phenomena in natural sciences: chaos, fractals, selforganization and disorder: concepts and tools*. Springer Science & Business Media.

64. Sornette, Davy, and Sornette (1990), Growth of fractal fault patterns., *Phys Rev Lett*, *65*(18), 2266–2269, doi:10.1103/physrevlett.65.2266.

65. Spassiani, I., & Sebastiani, G. (2016). Magnitude-dependent epidemic-type aftershock sequences model for earthquakes. *Physical Review E*, *93*(4), 042134.

66. Stanley, H. E. (1971). *Phase transitions and critical phenomena* (p. 7). Clarendon Press, Oxford.

67. Stram, D. O., & Lee, J. W. (1994). Variance components testing in the longitudinal mixed effects model. *Biometrics*, 1171-1177.

68. Veen, A., & Schoenberg, F. P. (2008). Estimation of space–time branching process models in seismology using an EM–type algorithm. *Journal of the American Statistical Association*, *103*(482), 614-624.

69. Vere-Jones, D. (2005). A class of self-similar random measure. *Advances in applied probability*, *37*(4), 908-914.

70. Wang, Q., Jackson, D., & Zhuang, J. (2010). Missing links in earthquake clustering models. *Geophysical Research Letters*, *37*(21), n/a–n/a. doi:10.1029/2010GL044858

71. Wang, Y., G. Ouillon, J. Woessner, D. Sornette, *and* S. Husen *(*2013*),* Automatic reconstruction of fault networks from seismicity catalogs including location uncertainty, *J. Geophys. Res. Solid Earth*, 118*,* 5956–5975*, doi:*10.1002/2013JB010164*.*

72. Wells, D. L., & Coppersmith, K. J. (1994). New empirical relationships among magnitude, rupture length, rupture width, rupture area, and surface displacement. *Bulletin of the seismological Society of America*, *84*(4), 974-1002.

73. Werner, M., A. Helmstetter, D. Jackson, and Y. Kagan (2011), High-Resolution Long-Term and Short-Term Earthquake Forecasts for CaliforniaHigh-Resolution Long-Term and Short-Term Earthquake Forecasts for California, *B Seismol Soc Am*, *101*(4), 1630–1648, doi:10.1785/0120090340.

74. Zaliapin, I., & Ben-Zion, Y. (2013). Earthquake clusters in southern California I: Identification and stability. *Journal of Geophysical Research: Solid Earth*, *118*(6), 2847-2864.

75. Zaliapin, I., & Ben-Zion, Y. (2016). A global classification and characterization of earthquake clusters. *Geophysical Journal International*, *207*(1), 608-634.

76. Zechar, J., Schorlemmer, D., Werner, M., Gerstenberger, M., Rhoades, D., & Jordan, T. (2013). Regional Earthquake Likelihood Models I: First-Order Results. *Bulletin of the Seismological Society of America*, *103*(2A), 787–798. doi:10.1785/0120120186

77. Zhuang, J., Ogata, Y., & Vere-Jones, D. (2004). Analyzing earthquake clustering features by using stochastic reconstruction. Journal of Geophysical Research: Solid Earth, 109(B5).

78. Zhuang, J., Y. Ogata, and D. Vere-Jones (2002), Stochastic Declustering of Space-Time Earthquake Occurrences, J Am Stat Assoc, 97(458), 369–380, doi:10.1198/016214502760046925

**Table 1:** Parameter estimates for the five models (see Section 2), which are variants of the gV-ETAS model, obtained by calibration on the Californian catalog using the EM algorithm; the maximum log-likelihood score obtained for each model is also reported in the first column.

| gV-ETAS variants (likelihood) | $\mu$ $\times 10^{-7}$ (day$^{-1}$km$^{-2}$) | $K$ | $a$ | $c$ $\times 10^{-2}$ (day) | $\omega$ | $d$ (km$^2$) | $\rho$ | $\gamma$ | $\beta_b$ | $\beta_a$ | $\delta$ |
|---|---|---|---|---|---|---|---|---|---|---|---|
| 1 (-60,917) | 2.31 | 0.58 | 1.01 | 1.74 | 0.17 | 0.24 | 0.53 | 1.11 | 2.38 | $= \beta_b$ $= 2.38$ | $\delta = 0$ |
| 2 (-60,909) | 2.31 | 0.58 | 1.01 | 1.74 | 0.17 | 0.24 | 0.53 | 1.11 | 2.48 | 2.36 | $\delta = 0$ |
| 3 (-60,280) | 2.30 | 0.58 | 1.00 | 1.73 | 0.17 | 0.24 | 0.53 | 1.11 | 2.36 | $= \beta_b$ $= 2.36$ | 0.75 |
| 4 (-60,349) | 2.28 | 0.58 | 1.00 | 1.73 | 0.17 | 0.24 | 0.53 | 1.12 | $= \beta_a + \delta$ $= 2.92$ | 2.28 | 0.64 |
| 5 (-60,266) | 2.31 | 0.58 | 1.00 | 1.73 | 0.17 | 0.24 | 0.53 | 1.11 | 2.46 | 2.35 | 0.74 |

**Table 2:** Maximum likelihood estimates of the parameters of the five models described in the text for the magnitude distribution of background and triggered earthquakes and their corresponding maximum log likelihood scores. The sixth column gives the p-values $p_{H_1}^{H_i}$ of Model *i* against the null hypothesis that Model 1 (ETAS) is the true model. Small values of these p-values reject the null in favor of all alternative models. The last column gives the p-values $p_{H_j}^{H_5}$ of the $j^{th}$ model taken as the null against Model 5. All models 1-4 are strongly rejected in favour of Model 5, with a confidence level better than 99.95%.

| Model | $\beta_b$ | $\beta_a$ | $\delta$ | MLL | $p_{H_1}^{H_i}$ | $p_{H_j}^{H_5}$ |
|---|---|---|---|---|---|---|
| 1 | 2.38 | 2.38 | 0 | -60,917 | | $2.6 \times 10^{-66}$ |
| **2** | 2.43 | 2.37 | 0 | -60,914 | $1.4 \times 10^{-2}$ | $2.5 \times 10^{-66}$ |
| **3** | 2.34 | 2.34 | 0.35 | -60,772 | $5.0 \times 10^{-65}$ | $5.3 \times 10^{-4}$ |
| **4** | 2.62 | 2.31 | 0.31 | -60,779 | $5.6 \times 10^{-62}$ | $3.4 \times 10^{-7}$ |
| **5** | 2.43 | 2.33 | 0.35 | -60,766 | $2.6 \times 10^{-66}$ | |

# Figures

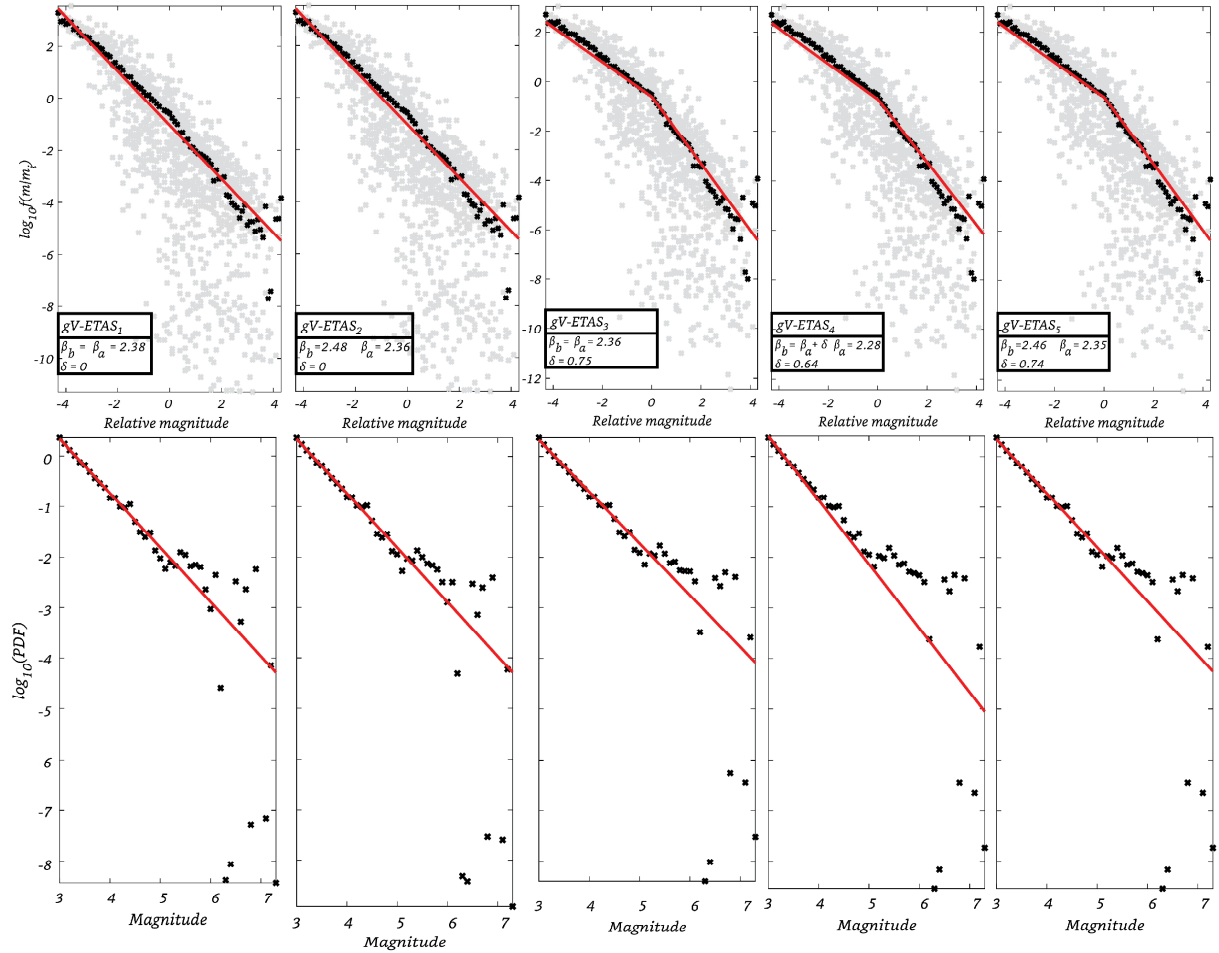

**Figure 1:** **(Upper panels)** grey crosses show the empirical PDFs of relative magnitudes of direct aftershocks for the 43 mainshock magnitude bins: [3, 3.1], [3.1,3.2]...[7.2,7.3]; black crosses show the master empirical PDF of the relative magnitudes obtained by stacking the individual scaled PDFs in each of the relative magnitude bins; solid red line represents the scaled theoretical PDFs corresponding to each of the five models using $\beta_a$ and $\delta$ parameters values shown in Table 1. See Section 4.1 for details of the construction of the figure. **(Bottom panels)** black crosses show the empirical PDFs of the background earthquakes; solid red lines stand for the theoretical PDF of background earthquakes, using values of parameters $\beta_b$ reported in Table 1.

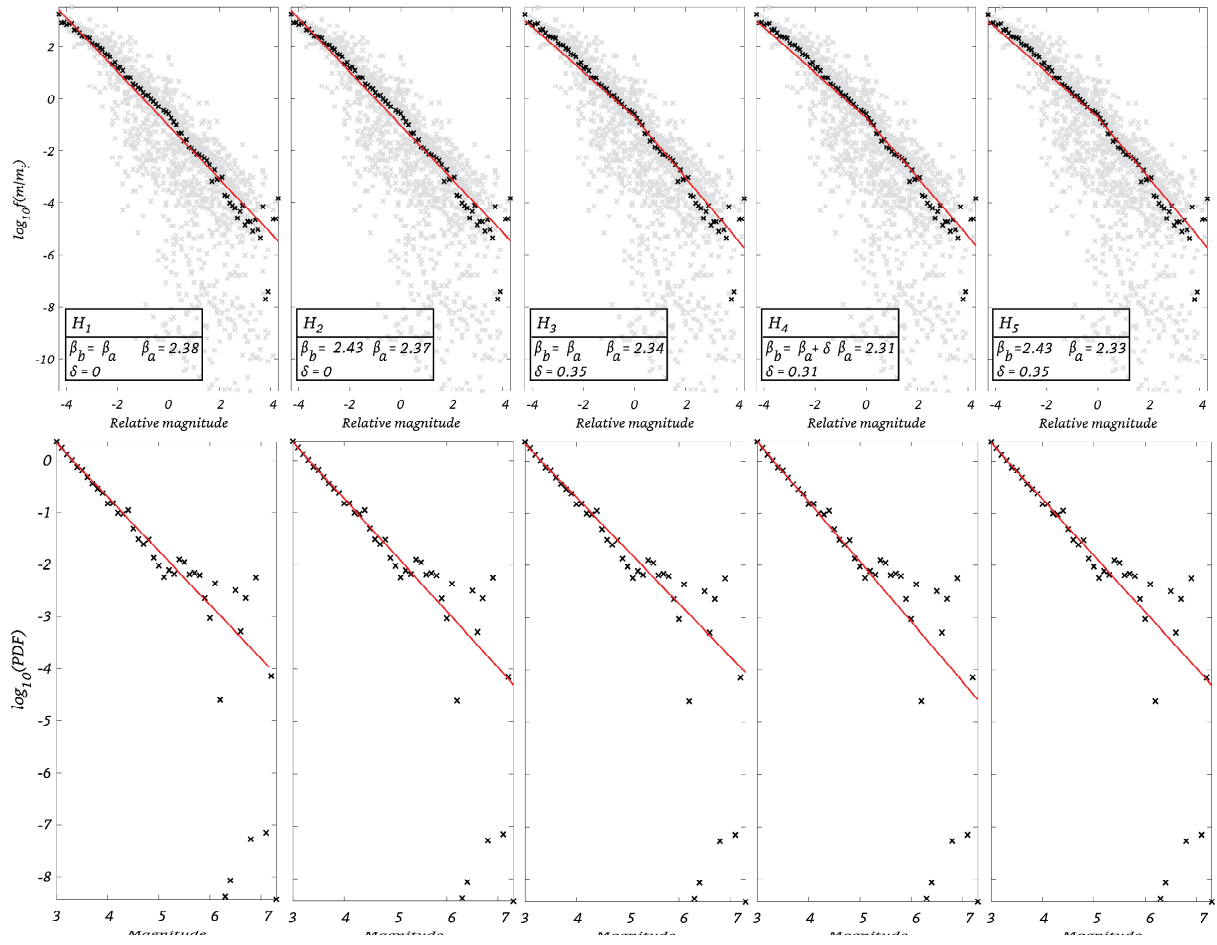

**Figure 2:** Same as Figure 1 but using the parameters of Table 2.

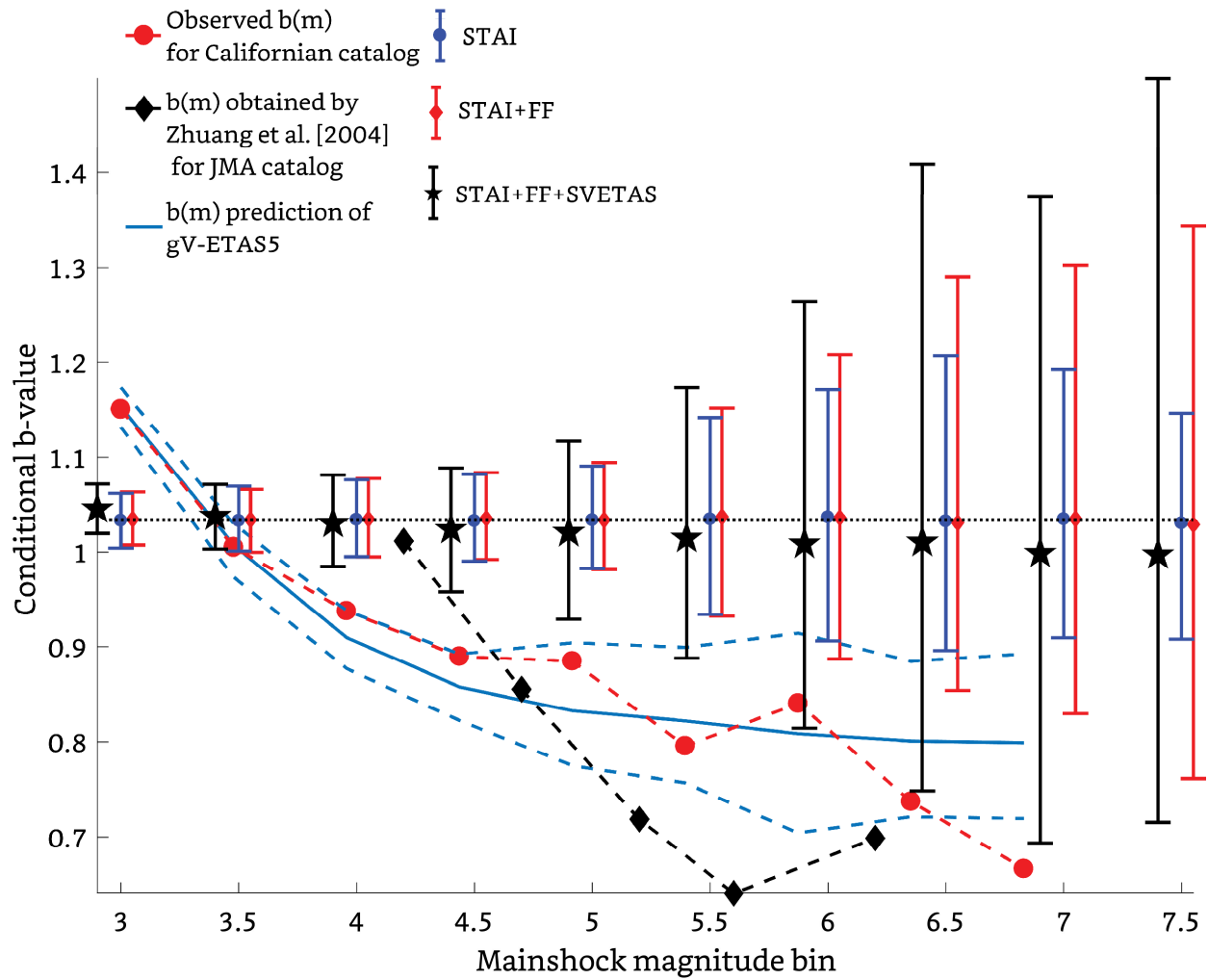

**Figure 3:** Median, 2.5%ile and 97.5%ile of the estimated apparent b-values of the GR law conditioned on mainshock magnitude on 1000 synthetic catalogues generated using Model 5 (gV-ETAS5) are shown using solid and dashed blue lines; Estimates of the same conditional b-values for the real Californian catalog are shown using red circles; Conditional b-values reported in [Zhuang et al., 2004] for the JMA catalog are shown using black diamonds; conditional b-values predicted by the basic ETAS (Model 1, gV-ETAS1) are shown using the solid horizontal black line; Median of the estimated conditional b-values for the catalogues simulated using STAI, STAI+FF and STAI+FF+SVETAS simulations are shown using blue circles, red diamonds and black stars respectively; 2.5%ile and 97.5%ile of the estimated conditional b-values for each of these simulations are shown using the error bars with same colors as the markers.

Supporting Information for

# Magnitude Of Earthquakes Controls The Size Distribution Of Their Triggered Events

Contents of this file

## Text S1: Standard ETAS model

In the standard Epidemic Type Aftershock Sequence (ETAS) model [Ogata, 1988], the conditional seismicity rate of magnitude m events, $\lambda(t, x, y, m | \mathcal{H}_t)$, at any location $(x, y)$ and time $t$, depends on the history of the earthquake occurrences up to t and is given by:

$$\lambda(t, x, y, m | \mathcal{H}_t) = \left[ \mu(x, y) + \sum_{i:t_i < t} g(t - t_i, x - x_i, y - y_i, m_i) \right] \beta e^{-\beta(m - M_0)} \qquad (S1)$$

In Equation (S1), $\mu(x, y)$ is the background intensity function, which is assumed to be independent of time, while $\mathcal{H}_t = \{(t_i, x_i, y_i, m_i): t_i < t\}$ represents the history of the process up to time $t$. The variables $(t_i, x_i, y_i, m_i)$ respectively correspond to the time, x-coordinate, y-coordinate and magnitude of the $i^{th}$ earthquake in the catalog, while $g(t - t_i, x - x_i, y - y_i, m_i)$ is the triggering function, defined in Equation (S2), quantifying how each past event influences earthquake occurrence in the future (see [Nandan et al., 2017] for the intuitive explanation of the different parameters):

$$g(x, y, t - t_i, x - x_i, y - y_i, m_i) =$$
$$K e^{a(m_i - M_0)} \frac{\omega c^\omega}{\{t - t_i + c\}^{1+\omega}} \frac{\rho d^\rho e^{\gamma\rho(m_i - M_0)}}{\pi\{(x - x_i)^2 + (y - y_i)^2 + d e^{\gamma(m_i - M_0)}\}^{1+\rho}} \qquad (S2)$$

$M_0$ is the magnitude of the smallest event able to trigger other earthquakes. The standard ETAS model defined above combines four robust empirical observations: the Omori law, which quantifies the decay rate of aftershocks following an event that occurred at time $t_i$; the spatial decay law, which quantifies how those aftershocks are distributed in space around the mainshock located at $(x_i, y_i)$; the Gutenberg-Richter (GR) law of the frequency of earthquake magnitudes [Gutenberg and Richter, 1944], assumed independent of seismicity rates; and the productivity law with exponent a, which quantifies the expected number of aftershocks that are directly triggered by an earthquake of magnitude $m_i$. In return, ETAS-based catalog declustering [Zhuang et al.,2004; Marsan and Lengliné, 2008; Veen and Schoenberg, 2008] allows one to compute, for a given event, the probability that it has been triggered by any of the previous ones, or that it is a background event.

Text S2: Additional log-likelihood contribution due to conditional magnitude distributions.

While obtaining the parameters of the gV-ETAS models, the following log-likelihood terms need to be added in the expected data log-likelihood defined in Equation (6) Section 2.2 in the main text). The log-likelihood terms are listed below in the same order as the list of five gV-ETAS models in Section 2.1 of the main text:

$$LL_1 = N \log \beta - \beta \sum_j (m_j - M_0) \qquad (S3)$$

$$LL_2 = N \log \beta_a + \left(\log \frac{\beta_b}{\beta_a}\right) \sum_j IP_j + (\beta_a - \beta_b) \sum_j IP_j \times (m_j - M_0) - \beta_a \sum_j (m_j - M_0) \qquad (S4)$$

$$LL_3 = \log \beta \sum_j IP_j - \beta \sum_j IP_j (m_j - M_0) + \sum_{ij} P_{ij} \log f_a(m_j | m_i, \beta, \delta) \qquad (S5)$$

$$LL_4 = \log(\beta_a + \delta) \sum_j IP_j - (\beta_a + \delta) \sum_j IP_j (m_j - M_0) + \sum_{ij} P_{ij} \log f_a(m_j | m_i, \beta_a, \delta) \qquad (S6)$$

$$LL_5 = \log \beta_b \sum_j IP_j - \beta_b \sum_j IP_j (m_j - M_0) + \sum_{ij} P_{ij} \log f_a(m_j | m_i, \beta_a, \delta) \qquad (S7)$$

where, $\log f_a(m | m_i, \beta, \delta) =$
$$\begin{cases} -\log\left[\frac{2\delta}{\beta^2 - \delta^2}\left\{\left(\frac{\beta + \delta}{2\delta}\right) e^{-(\beta - \delta)M_0} - e^{-(\beta - \delta)m_i}\right\}\right] - (\beta - \delta)m & \forall m \le m_i \\ -\log\left[\frac{2\delta}{\beta^2 - \delta^2}\left\{\left(\frac{\beta + \delta}{2\delta}\right) e^{-(\beta - \delta)M_0} - e^{-(\beta - \delta)m_i}\right\}\right] + 2\delta m_i - (\beta + \delta)m & \forall m > m_i \end{cases}$$

Note that, in the above equations, $N$ represents the total number of the earthquakes present in the catalog.

The above formulations of the log-likelihood have been defined assuming that the underlying magnitude distributions of the earthquakes are continuous in nature. While there is no empirical evidence to doubt it, reported magnitudes are often binned in the earthquake catalogs at intervals of 0.1 units. The ANSS catalog is no exception. Given that the magnitudes in the earthquake catalogs are discretized in $\Delta m$ units, the log likelihoods can be reformulated as follows:

$$LL_1^{\Delta m} = N \log(1 - e^{-\beta \Delta m}) - \beta \sum_j (m_j - M_0) \qquad (S8)$$

$$LL_2^{\Delta m} = N \log(1 - e^{-\beta_a \Delta m}) + \log\left(\frac{1 - e^{-\beta_b \Delta m}}{1 - e^{-\beta_a \Delta m}}\right) \sum_j IP_j +$$

$$(\beta_a - \beta_b) \sum_j IP_j \times (m_j - M_0) - \beta_a \sum_j (m_j - M_0) \qquad (S9)$$

$$LL_3^{\Delta m} = \log\left(1 - e^{-\beta\Delta m}\right)\sum_j IP_j - \beta\sum_j IP_j(m_j - M_0) +$$

$$\sum_{ij} P_{ij}\log f_a(m_j|m_i, \beta, \delta, \Delta m) \qquad (S10)$$

$$LL_4^{\Delta m} = \log\left(1 - e^{-\beta_b\Delta m}\right)\sum_j IP_j - \beta_b\sum_j IP_j(m_j - M_0) +$$

$$\sum_{ij} P_{ij}\log f_a(m_j|m_i, \beta_a, \delta, \Delta m) \qquad (S11)$$

$$LL_5^{\Delta m} = \log\left(1 - e^{-(\beta_a+\delta)\Delta m}\right)\sum_j IP_j - (\beta_a + \delta)\sum_j IP_j(m_j - M_0) +$$

$$\sum_{ij} P_{ij}\log f_a(m_j|m_i, \beta_a, \delta, \Delta m) \qquad (S12)$$

where,

$$\log f_a(m|m_i, \beta, \delta, \Delta m)$$

$$= \begin{cases} (\beta - \delta)(m_i - m) + (\beta + \delta)\Delta m + \log\left\{\dfrac{\left(1 - e^{-(\beta+\delta)\Delta m}\right)\left(1 - e^{-(\beta-\delta)\Delta m}\right)}{e^{-(\beta-\delta)M_0} - e^{-(\beta-\delta)(m_i+\Delta m)}}\right\} & \forall m \leq m_i \\[4mm] (\beta + \delta)(m_i - m) + (\beta + \delta)\Delta m + \log\left\{\dfrac{\left(1 - e^{-(\beta+\delta)\Delta m}\right)\left(1 - e^{-(\beta-\delta)\Delta m}\right)}{e^{-(\beta-\delta)M_0} - e^{-(\beta-\delta)(m_i+\Delta m)}}\right\} & \forall m > m_i \end{cases}$$

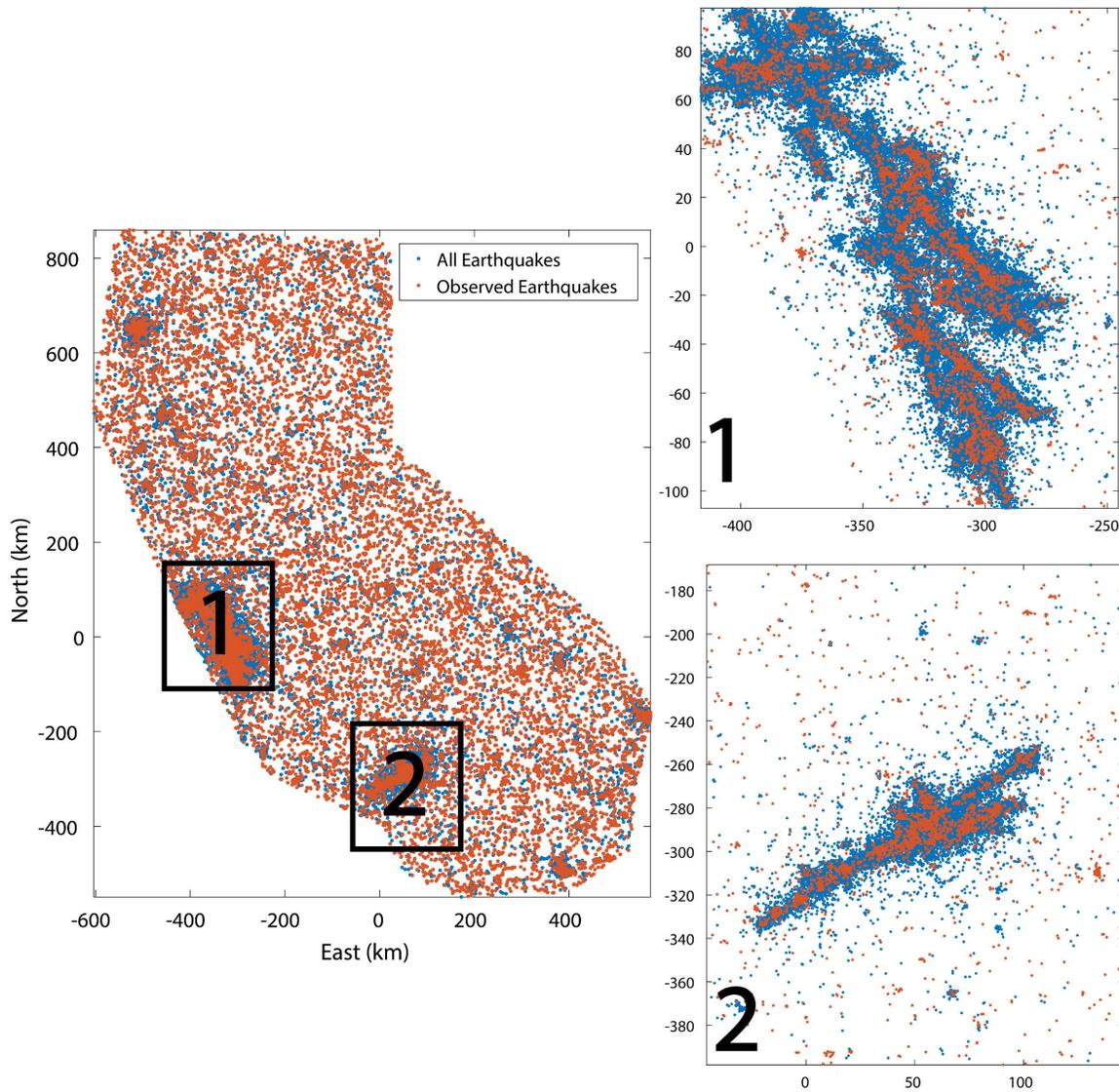

Figure S2: An example spatial distribution of all (blue dots) and observed (red dots) earthquakes ($M \geq 3$) obtained in the hybrid of STAI and FF simulation.

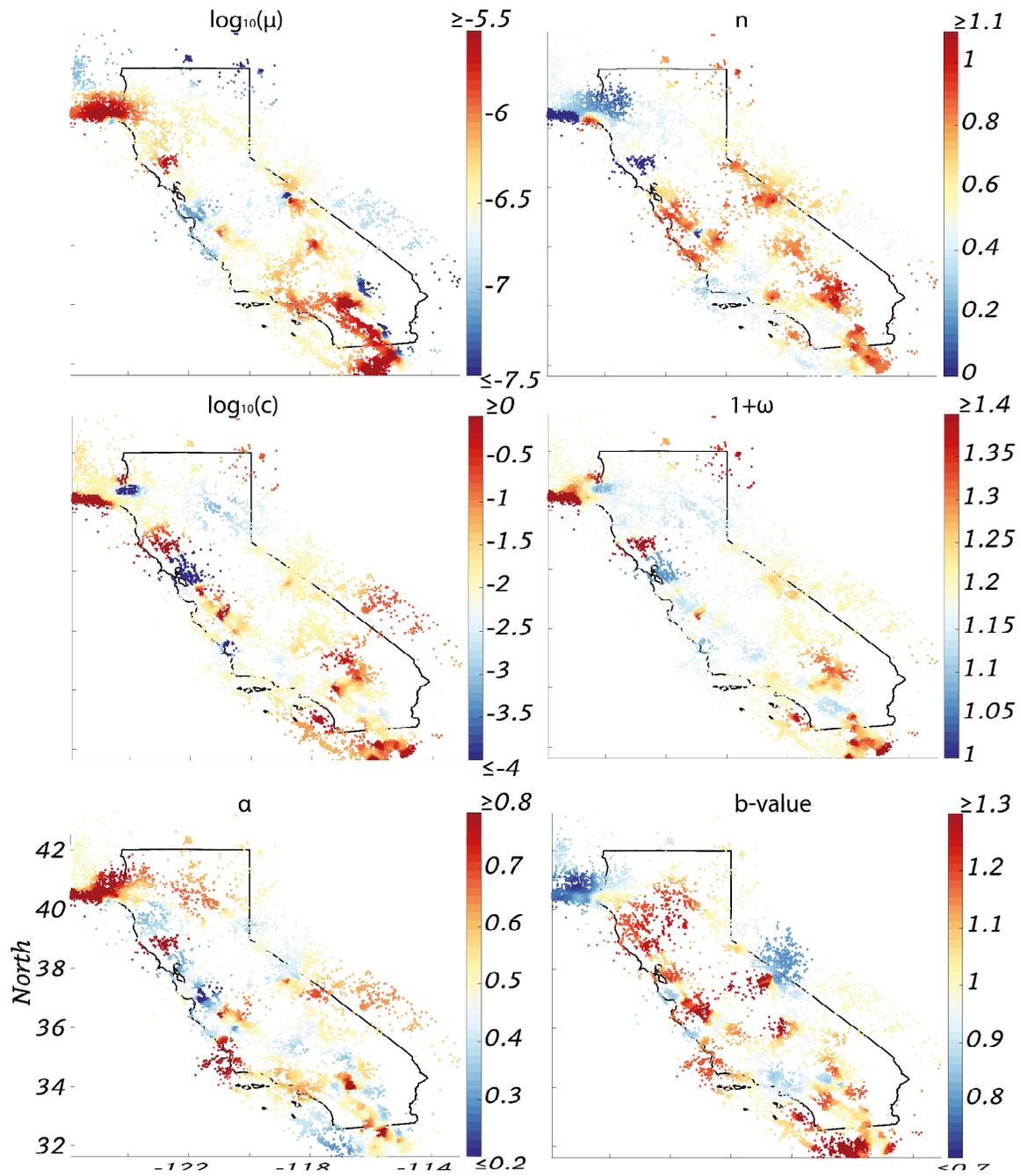

Figure S2: Ensemble estimated of the spatially variable ETAS parameters $\left(\boldsymbol{\mu}, \boldsymbol{n}, \boldsymbol{c}, \boldsymbol{\omega}, \boldsymbol{\alpha} = \frac{a}{log(10)}, \boldsymbol{b-value} = \frac{\beta}{log(10)}\right)$ at the location of the earthquakes used to calibrate the model.